\documentclass[twocolumn,aps,prc,superscpologyNet,superscriptaddress,riptaddress,amsmath,amssymb]{revtex4-1}
\usepackage{epsfig,dsfont,amssymb,amsmath,amsthm,amsfonts,amsbsy,mathrsfs}
\usepackage{graphicx}
\usepackage{amsmath}
\usepackage{amssymb}%
\usepackage{booktabs}
\usepackage{multirow}
\usepackage{lineno}
\usepackage{dcolumn}% Align table columns on decimal point
\usepackage{bm}% bold math
\usepackage{color}
\usepackage{float}
\usepackage{graphicx}
\usepackage[colorlinks,
           bookmarks=true,
           linkcolor=blue,
           urlcolor=blue,
            anchorcolor=black,
            citecolor=blue
            ]{hyperref}
\usepackage[all]{hypcap}

\begin{document}

\title{Identifying weak critical fluctuations of intermittency in heavy-ion collisions with topological machine learning}

\author{Rui Wang}
\affiliation{Key Laboratory of Quark and Lepton Physics (MOE) and Institute of Particle Physics, \\
Central China Normal University, Wuhan 430079, China}

\author{Chengrui Qiu}
\affiliation{Key Laboratory of Quark and Lepton Physics (MOE) and Institute of Particle Physics, \\
Central China Normal University, Wuhan 430079, China}

\author{Chuan-Shen Hu}
\affiliation{Division of Mathematical Sciences, School of Physical and Mathematical Sciences, \\
Nanyang Technological University, Singapore, 637371, Singapore}

\author{Zhiming Li}
\email{lizm@mail.ccnu.edu.cn}
\affiliation{Key Laboratory of Quark and Lepton Physics (MOE) and Institute of Particle Physics, \\
Central China Normal University, Wuhan 430079, China}
\affiliation{College of Physics and Electronic Engineering, Hanjiang Normal University, Shiyan 442000, China}

\author{Yuanfang Wu}
\affiliation{Key Laboratory of Quark and Lepton Physics (MOE) and Institute of Particle Physics, \\
Central China Normal University, Wuhan 430079, China}

\begin{abstract}
Large density fluctuations of conserved charges have been proposed as a promising signature for exploring the QCD critical point in heavy-ion collisions. These fluctuations are expected to exhibit a fractal or scale-invariant behavior, which can be probed by intermittency analysis. Recent high-energy experimental studies reveal that the signal of critical fluctuations related to intermittency is very weak and thus could be easily obscured by the overwhelming background particles in the data sample. Employing a point cloud neural network with topological machine learning, we can successfully classify weak signal events from background noise by the extracted distinct topological features, and accurately determine the intermittency index for weak signal event samples.
\end{abstract}

\maketitle
%%%%%%%%%%%%%%%%%%%%%%%%%%%%%%%%%%%%%%%%%%%%%%%%%%%%%%%%%%%%%%%%%%%%%%%%%%%%%%%
\section{Introduction}\label{sec:Introduction}
%%%%%%%%%%%%%%%%%%%%%%%%%%%%%%%%%%%%%%%%%%%%%%%%%%%%%%%%%%%%%%%%%%%%%%%%%%%%%%%
The Quantum Chromodynamics (QCD) phase diagram maps the phases of strongly interacting matter under varying conditions of temperature and baryon chemical potential~\cite{StephanovPD,adams2005experimental,conservecharge0,conservecharge1,Lattice1,Lattice2}.  It includes phases such as hadronic matter, where quarks are held together within hadrons by the strong force, and the quark-gluon plasma, a state where quarks and gluons are deconfined. It is suggested that at low baryon chemical potential ($\mu_B$) and high temperature ($T$), the transition is a smooth crossover, while at low $T$ and high $\mu_B$, the phase transition could be of the first-order. A critical point (CP) in the QCD phase diagram represents the endpoint of the first-order phase transition line between hadronic matter and quark-gluon plasma~\cite{CEP1,CEP2,CEP3}. Studying the CP is of great importance as it offers deep insights into the properties of QCD matter, validates theoretical models, and helps us understand the evolution of the universe shortly after the Big Bang. Experimental efforts to locate the CP, such as the Beam Energy Scan (BES) program at the Relativistic Heavy-ion Collider (RHIC)~\cite{RHIC1, RHIC2}, are crucial for advancing our knowledge in high-energy nuclear physics and for testing the predictions of QCD under extreme conditions.

It is suggested that in high-energy collisions, event-by-event fluctuations of conserved charges could provide a distinctive signature for the phase transition in the QCD phase diagram~\cite{fluctuation1,fluctuation2,fluctuation4,fluctuation5}. The singularity at the CP, where the transition is expected to be of the second-order, may lead to significant density fluctuations in momentum space~\cite{StephanovPD,baryon_density_fluctuations2}. These fluctuations can be measured using scaled factorial moment (SFM) and are anticipated to exhibit a clear intermittency behavior, which is characterized by a fractal pattern of power-law or self-similar nature in the distribution of final-state particles in heavy-ion collisions~\cite{intermittency, CMC1, CMC2,CMC3,CMC4,CMC5,critical_fluctuation1,critical_fluctuation2}.

In high-energy experimental studies, intermittency was first observed in central Si + Si collisions at the maximum SPS energy of 158$A$ GeV by the NA49 collaboration~\cite{NA49}. However, the NA61/SHINE experiment at near SPS energies did not detect any intermittency signals at either 150$A$ GeV~\cite{NA61_150} or 13A-75$A$ GeV~\cite{NA61} in central Ar + Sc collisions. In the meantime, the STAR experiment at RHIC energies~\cite{STARBESI} observed a power-law intermittency behavior of SFM ratios in central Au + Au collisions. The extracted scaling exponent displayed a non-monotonic energy dependence over the center-of-mass energy range $\sqrt{s_{NN}}=7.7-200 \enspace {\rm GeV}$. Further investigations reveal that the intermittency observed in NA49 can be reproduced by a mixed sample with 99\% background random tracks and 1\% signal particles generated from a Critical Monte Carlo (CMC) model~\cite{NA49}. For NA61 data, the upper limit on the signal particle fraction is around 1\%~\cite{NA61_150,NA61}. The STAR experimental results are consistent with a mix of approximately 1-2\% signal particles~\cite{critical_fluctuation2}. These findings indicate that even if critical fluctuations of intermittency exist in heavy-ion collisions, the signal is very weak, consisting in only a few percent of the whole sample, making it easy to be overshadowed by experimental noise. Therefore, before comparing and understanding the various experimental measurements on intermittency, it is necessary to identify and extract the weak signal from the predominant noisy background.

With the advancement of modern computer hardware and artificial intelligence, machine learning (ML)~\cite{ML1Science,ML2Nature}, as a data-driven approach, provides prospective opportunities for this investigation. State-of-the-art ML utilizes sophisticated algorithms to analyze complex datasets and identify subtle patterns that traditional methods might miss. Deep learning techniques have been effectively applied to particle collision data analysis, helping to identify rare events indicative of new physics phenomena~\cite{ML_Apply}. A combination of dynamical edge convolution and point cloud neural networks demonstrates robust pattern recognition capabilities in identifying self-similarity from uncorrelated backgrounds~\cite{ML_mix}. Point cloud networks~\cite{ML_mix,DTFE1,pointcloud1,pointcloud2} are often employed in these studies due to their ability to effectively handle and analyze the complex and high-dimensional data obtained from particle detectors in high-energy experiments. 

In the meantime, geometry and topology have emerged as powerful tools for identifying phase transitions in complex systems~\cite{XY_Model,PHPC1,PHPC2,PHPC3,TDAPC}. These approaches provide deeper insights beyond traditional thermodynamic measurements. By analyzing the shape and connectivity of a system using tools from topological data analysis (TDA), one can detect subtle changes that signify transitions between different phases. Techniques such as persistent homology (PH) are used to analyze the characteristic topological features of the data sample, helping to identify the critical point and the nature of phase transitions. 

Recent studies suggest that topological machine learning, a methodology that combines TDA with ML, can enhance the performance of point cloud neural networks~\cite{TML1,TML2,TML3}. In this method, TDA provides deep learning models with a more comprehensive understanding of the underlying structure in the data, while deep learning offers TDA powerful tools for learning complex relationships between the shape and properties of the data. Point cloud neural networks utilize TDA to comprehend data through shape or local features, significantly reducing training time and enhancing capabilities such as pattern recognition. Furthermore, the interpretability and transferability aspects of ML are closely intertwined with TDA~\cite{TML}. In this context, PH has been proposed as one of the foremost techniques for nonparametrically identifying important topological characteristics of a given point cloud. PH computes the frequency and significance of topological features by constructing a sequence of geometric complexes~\cite{PH_features1,PH_features2,PH_features3}.  It has been used to describe topological invariants of data and is capable of quantifying features such as clusters, holes, and voids within the data~\cite{TDA_Introduction}. This methodology has been applied in identifying long-range flow correlations in nuclear collisions~\cite{DTFE1}, describing and evolving cosmic structures~\cite{applycosmic1,applycosmic2,applycosmic3}, probing universal dynamics in a gluonic plasma~\cite{pointandtopo}, and discerning phase transitions in the mean-field XY and lattice models~\cite{XY_Model}.

In this study, we incorporate persistent homology and machine learning to explore the significant topological characteristics associated with the critical fluctuations of intermittency in heavy-ion collisions. We construct a TopoPointNet, an innovative framework that integrates TDA with point cloud networks, to classify weak intermittency signals from a massive amount of background noise. The paper is organized as follows:  In Sec.~\ref{sec:event}, we introduce the intermittency analysis and the data samples generated by the CMC model and the Ultra-relativistic Quantum Molecular Dynamics (UrQMD) model. A brief description of TDA and the TopoPointNet architecture are provided in Sec.~\ref{sec:net}. The results of identifying and extracting weak intermittency signals from the background are presented and discussed in Sec.~\ref{sec:result}. Finally, we give a short summary and outlook of the work in Sec.~\ref{sec:summary}.

%%%%%%%%%%%%%%%%%%%%%%%%%%%%%%%%%%%%%%%%%%%%%%%%%%%%%%%%%%%%%%%%%%%%%%%%%%%%%%%
\section{Intermittency analysis and data samples}\label{sec:event}
%%%%%%%%%%%%%%%%%%%%%%%%%%%%%%%%%%%%%%%%%%%%%%%%%%%%%%%%%%%%%%%%%%%%%%%%%%%%%%%
Intermittency refers to the phenomenon where fluctuations in particle densities exhibit power-law behavior when observed at different scales. It provides insights into the underlying dynamics of heavy-ion collisions, which may indicate critical phenomena and phase transitions in nuclear matter. Intermittency can be studied by calculating the scaled factorial moment~\cite{CMC1,intermittency_and_fractal_dimension1,intermittency_and_fractal_dimension2,intermittency_and_fractal_dimension3}, a statistical measure used to analyze the distribution of produced particles. In a $D$-dimensional momentum space, the $q$th-order SFM is defined as:
\begin{equation}
F_{q}(M)=\frac{\langle\frac{1}{M^{D}}\sum_{i=1}^{M^{D}}n_{i}(n_{i}-1)\cdots(n_{i}-q+1)\rangle}{\langle\frac{1}{M^{D}}\sum_{i=1}^{M^{D}}n_{i}\rangle^{q}}.
 \label{Eq:FM}
\end{equation}
\noindent Here, $M^D$ is the number of cells into which the phase space is divided with equal size in momentum space, $n_{i}$ is the measured number of particles in the $i$th cell, and the angular brackets indicate an average over the entire event sample.

As the system approaches the critical point, fluctuations in particle densities become more pronounced. This increased fluctuation can be captured by the SFM, which exhibits a power-law behavior with respect to the number of cells~\cite{CMC1,CMC4}:
\begin{equation}
F_{q}(M)\sim (M^{D})^{\phi_{q}}.
 \label{Eq:PowerLaw}
\end{equation}
\noindent Where $\phi_{q}$  is the intermittency index, specifying the strength of the intermittency behavior. This index provides information about the nature of the fluctuations and correlations within the system. It is argued that the QCD critical point, if it exists, interacts with a zero-mass scalar field ($\sigma$-field) that may reach the two-pion ($\pi^{+}\pi^{-}$ pairs) threshold at lower temperatures~\cite{intermittencyindex,intermittencyindex-2,intermittencyindex-3,intermittencyindex-4}. The $\sigma$ condensate describes a quantum state linked to the chiral condensate, where the isoscalar $\sigma$-field embodies the quantum numbers and critical properties of the condensate, playing a crucial role in understanding the characteristics of the critical point. If critical fluctuations persist through the evolution of heavy-ion collisions, the second-order intermittency index $\phi_{2}$ is predicted to be $2/3$ for $\sigma$ condensate based on calculations from a critical equation-of-state belonging to the 3D Ising universality class~\cite{intermittencyindex}.

To simulate critical fluctuations related to self-similar intermittency, the CMC model~\cite{CMC1, CMC2,CMC3,CMC4,CMC5} is used to generate event samples that include information on particle momenta and densities. These samples can then be used to compute SFMs and study the scaling behavior of intermittency. In this model, momentum distributions of final-state particles are generated using the Lévy random walk algorithm~\cite{Levy_random}, which requires the probability density $\rho(p)$ between two adjacent walks to follow
\begin{equation}
\rho(p)=\frac{\mu p_{\rm min}^{\mu}}{1-(p_{\rm min}/p_{\rm max})^{\mu}}p^{-1-\mu}.
 \label{Eq:probLevy}
\end{equation}
\noindent Where $\mu$ is the Lévy exponent associated with the intermittency index, $p$ refers to the relative momentum between two neighboring particles, and $p_{\rm min}$ and $p_{\rm max}$ are the minimum and maximum values of $p$, respectively. The model parameters in Eq.~\eqref{Eq:probLevy} are set to be $\mu = 1/6$ and $p_{\rm min}/p_{\rm max} = 10^{-7}$ for a critical system belonging to the 3D Ising universality class~\cite{intermittency_and_fractal_dimension1,intermittency_and_fractal_dimension2,intermittency_and_fractal_dimension3}. The CMC model serves as a powerful tool in the study of critical fluctuations of intermittency in high-energy physics, bridging the gap between theoretical predictions and experimental observations.

The UrQMD model is a microscopic framework designed to simulate and analyze the dynamical evolution of high-energy collisions~\cite{UrQMD1, UrQMD2}. This model integrates principles from quantum mechanics, special relativity, and molecular dynamics to offer detailed insights into the behavior of nuclear matter under extreme conditions. It includes relativistic kinematics and dynamics to ensure energy and momentum conservation. The UrQMD model accounts for the creation and annihilation of particles, as well as the scattering processes occurring during collisions, by tracking the interactions of individual hadrons and their propagation through space-time. The model has been widely and successfully applied in studies of $p + p$, $p + A$, and $A + A$ interactions in experiments across a broad range of collision energies, from a few GeV to the energies reached at CERN LHC. Results from the UrQMD model are often compared with experimental data from facilities like the RHIC~\cite{critical_fluctuation2, critical_fluctuation1, CMC5, CMC4} and the LHC~\cite{UrQMD3, UrQMD4}. These comparisons help validate the model and refine our understanding of heavy-ion collisions.

Current experimental intermittency measurements have shown that the weak signal, indicative of critical fluctuations in heavy-ion collisions, is only a few percent~\cite{NA49, NA61_150, NA61, critical_fluctuation2}. To simulate a weak signal event dataset, we mix the CMC particles containing intermittency fluctuations with pure UrQMD background particles. Signal events are obtained by replacing a portion of particles in the corresponding UrQMD events with CMC particles. We define the replacement ratio, $\lambda = N_{\rm CMC}/N_{\rm UrQMD}$, to represent the multiplicity ratio of CMC events to UrQMD events. This ratio indicates the strength with which the critical signal is embedded in the signal events. The detailed process for generating signal events is as follows:

\begin{enumerate}
    \item Configure the simulation collision parameters and generate a UrQMD event sample.
    \item Randomly select a particle in the UrQMD event as the initial particle to generate the corresponding CMC data sample.
    \item Randomly select one particle from the UrQMD event and another from the CMC dataset, ensuring the constraint $\left | p_T^{\rm CMC}-p_T^{\rm UrQMD} \right | < 0.2$ GeV/$c$. 
    \item Replace the selected UrQMD particle with the chosen CMC particle, and simultaneously remove the CMC particle from the CMC dataset.
    \item Repeat steps 3 to 4 until the replacement ratio $\lambda$ reaches the predetermined target value.
\end{enumerate}

By following this procedure, the transverse momentum spectra remain nearly unchanged between the signal event and the original UrQMD background event samples. The replacement ratio $\lambda$ is a small value according to experimental measurements.

In this study, we use the cascade UrQMD model (version 3.4) to generate background event samples in 0\%-5\% most central Au + Au collisions at $\sqrt{s_\mathrm{NN}}$ =19.6 GeV. We generate $1.28\times 10^5$ signal events along with an equal number of UrQMD events for $\lambda = 10\%$, and $2.56\times 10^5$ signal events and UrQMD events for $\lambda = 5\%$. We apply the same analysis techniques and kinematic cuts on the data samples as those used in the STAR experiment~\cite{STARBESI,STAR1}. Charged particles are selected within the pseudo-rapidity window ($|\eta| <0.5$) and transverse momentum interval ($0.2<p_{T}<2$ GeV/$c$) . The statistical uncertainty is estimated using the bootstrap method~\cite{RefBootstrap}.

%%%%%%%%%%%%%%%%%%%%%%%%%%%%%%%%%%%%%%%%%%%%%%%%%%%%%%%%%%%%%%%%%%%%%%%%%%%%%%%
\section{Topological machine learning}\label{sec:net}
%%%%%%%%%%%%%%%%%%%%%%%%%%%%%%%%%%%%%%%%%%%%%%%%%%%%%%%%%%%%%%%%%%%%%%%%%%%%%%%
\begin{figure*}
    \centering 
    \includegraphics[scale=0.8]{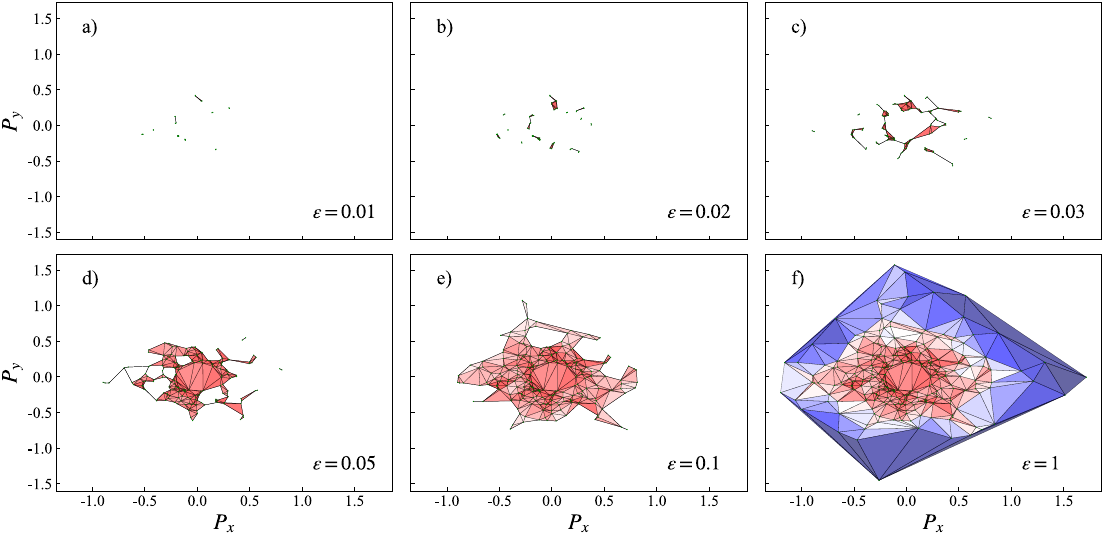}
    \caption{The Delaunay triangulation filtration process of persistent homology on a signal event sample with $\lambda$ = 5\% in two-dimensional momentum space. When the filtration level $\varepsilon$ is small, a large number of clusters emerge. As the filtration level increases, these clusters begin to connect, forming holes, and eventually coalesce into a single structure. Red color indicates areas of high density, while blue color represents regions of low density.}
    \label{Fig:filtration}
\end{figure*}

Topological machine learning, or TDA-based machine learning, is a powerful and versatile tool for modern data analysis, which merges geometric and topological insights with the predictive power of machine learning to more effectively tackle complex, high-dimensional problems \cite{TML1,TML2,TML3}. The topological method, which studies the connectivity of geometric objects, reveals inherent independent components, holes, and higher-dimensional cavities \cite{Topology}. With deep connections to theories such as homology, Morse theory, and size functions, the topological characteristics of an object remain invariant under shape deformations, demonstrating robustness in capturing global information despite perturbations and fluctuations. However, in conventional ML models, these higher-dimensional and invariant features are often overlooked during the design of ML architectures. TDA tools, particularly when assisted by multiscale filtrations, excel at detecting smaller and more subtle clusters and topological structures that traditional methods may fail to uncover. The filtration is a process of gradually building a nested sequence of simplicial complexes or networks from a given dataset, where each step in the filtration represents a different scale or level of connectivity.
    
Persistent homology, a core concept in TDA, is an effective tool in computational topology used to analyze the topological features of data sets, particularly by tracking how these features persist as filtering levels change continuously~\cite{PH_features1,PH_features2,PH_features3}. By employing homology theory from algebraic topology~\cite{cal_betti}, PH captures the persistence and changes of connected components, loops, and higher-dimensional voids within filtered geometric objects. In particular, a type of combinatorial geometric object known as a simplicial complex provides a systematic and computationally feasible method for constructing the underlying spaces used in computing PH, referred to as the filtration of simplicial complexes~\cite{PH_features1,PH_features2,PH_features3}. A simplicial complex consists of geometric simplices of various dimensions: vertices are treated as 0-simplices, edges as 1-simplices, triangles as 2-simplices, and so on, with no overlapping between simplices except where they share their boundary faces~\cite{cal_betti}. Our study of PH follows three key steps \cite{DTFE1, DTFE2}: first, we apply the Delaunay Triangulation Field Estimator (DTFE) to the given point cloud dataset; next, we construct the simplicial complex filtration and derive the PH from it; finally, we compute the topological features throughout the associated PH.

\paragraph{DTFE-based filtration construction}

In the process of computing PH, various methods can be used to construct the underlying filtrations of simplicial complexes, including Vietoris--Rips complexes~\cite{VRComplex}, Alpha complexes~\cite{AlphaComplex-1, AlphaComplex-2}, and $\check{\mathrm{C}}$ech complexes~\cite{TopologicalComplexes}. In this work, for a given set of points in the Euclidean space, we utilize Delaunay triangulation as the foundation for constructing the filtration of simplicial complexes~\cite{delaunay1,delaunay2}. Specifically, in $\mathbb{R}^2$ or $\mathbb{R}^3$, Delaunay triangulation divides the convex hull of the point set into triangles, also known as triangle meshes, by maximizing the minimum angles and ensuring that the circumcircle of each triangle contains no other points from the set. This process provides a unique and robust representation of the point set in $\mathbb{R}^2$ or $\mathbb{R}^3$. In particular, the Delaunay triangulation $D(X)$ of a dataset of points $X \subseteq \mathbb{R}^2$ (or $\mathbb{R}^3$) forms a simplicial complex, where the points in the dataset constitute the collection of all its $0$-simplices.

For a given point cloud dataset $X$, after the Delaunay triangulation, we define a distance field as a function $f: X \rightarrow \mathbb{R}$ using a field estimation method~\cite{distanceField}, which serves as the basis for constructing the filtration. Specifically, each point $x \in X$ is assigned as the geometric distance to its nearest neighbor. Compared to traditional methods for constructing density fields, our results demonstrate that distance fields exhibit greater sensitivity to intermittency signals, more effectively highlighting signals within a complex background. 

Subsequently, based on the Delaunay triangulation $D(X)$ and the distance field function $f: X \rightarrow \mathbb{R}$, we construct the sub-level set filtration, a classic method for building topological filtrations in point cloud and digital image data~\cite{SublevelFiltration-1, SublevelFiltration-2}. Mathematically, for every filtration level $\varepsilon$, the sub-level set $L_\varepsilon^+ = \left \{x \in X \ | \ f(x)\le{\varepsilon}\right \}$ collects data points with values less than or equal to $\varepsilon$. As the filtration level increases, the set of points is continually populated until it encompasses all points in the event i.e., $L_\varepsilon^+ \subseteq L_\delta^+$ whenever $\varepsilon \leq \delta$, and $L_\varepsilon^+ = X$  for sufficiently large $\varepsilon$. Then, the complex $K_{\varepsilon}$ formed for each $\varepsilon$ consists of the corresponding set of points, along with the line segments and triangles that can be formed from this set and are contained within the triangulation. Formally, $K_\varepsilon = \{ \sigma \in D(X) \ | \ {\rm vert}(\sigma) \subseteq L_\varepsilon^+ \}$, where ${\rm vert}(\sigma)$ denotes the set of vertices of the simplex  $\sigma$.  Consequently, by choosing an increasing sequence of filtration levels $0 = \varepsilon_0 < \varepsilon_1 < \varepsilon_2 < \cdots < \varepsilon_m$ with sufficiently large $\varepsilon_m$, the topological filtration
\begin{equation}
\label{Eq. built filtration of simplicial complexes}
\emptyset = K_{\varepsilon_0} \subseteq K_{\varepsilon_1} \subseteq K_{\varepsilon_2} \subseteq \cdots \subseteq K_{\varepsilon_m} = D(X)
\end{equation}
is obtained. 

Figure~\ref{Fig:filtration} illustrates how the complex evolves continuously as the filtration level $\varepsilon$ changes. As $\varepsilon$ increases, more simplices are added, causing changes in the topological information, such as connected components and loop structures, throughout the process. Fig.~\ref{Fig:filtration} (a)-(c) show the complexes during the initial stages of filtration corresponding to \(\varepsilon=0.01,\varepsilon=0.02,\varepsilon=0.03\), respectively. At these stages, a few points that meet the filtration levels begin to emerge, but the resulting components are primarily scattered. In the late stages of filtration, as shown in Fig.~\ref{Fig:filtration} (d)-(f), more points emerge, connecting the previously scattered components. During this process, holes may appear, ultimately resulting in a complete Delaunay triangulation structure.

\begin{figure*}
\centering 
\includegraphics[scale=0.6]{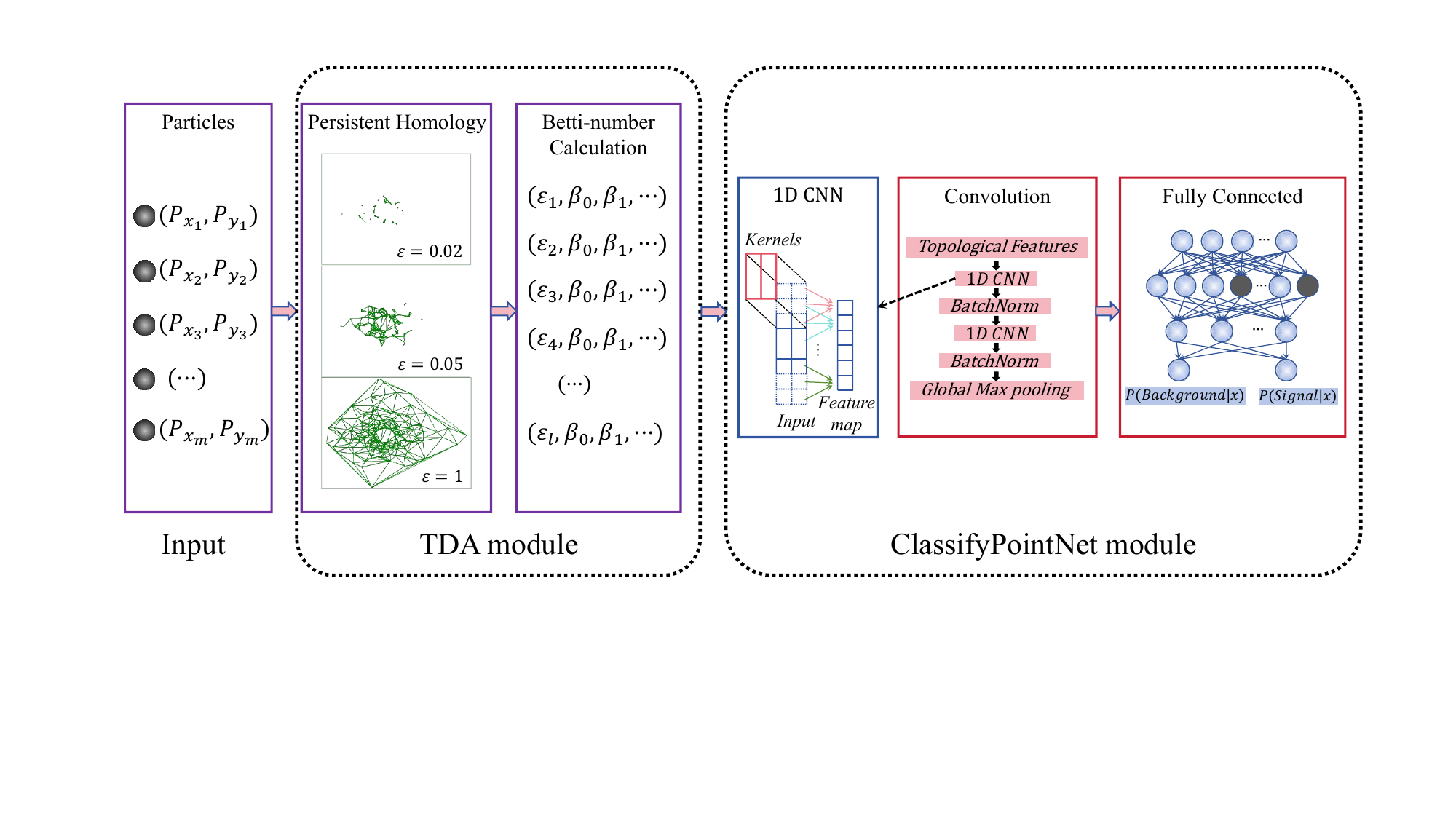}
\caption{The TopoPointNet architecture used in the analysis.}
\label{Fig:TopoNet}
\end{figure*}

\paragraph{PH computation}
Leveraging the homology of each simplicial complex, once the filtration of simplicial complexes is constructed, persistent homology can be computed. Specifically, for the filtration defined in Eq.~\eqref{Eq. built filtration of simplicial complexes} and a non-negative integer \( p \), the \( p \)-th PH of this filtration is defined as the following sequence of abelian groups and group homomorphisms:
\begin{equation}
\label{Eq. PH of filtration}
0 = H_p(K_{\varepsilon_0}) \xrightarrow{f_0} H_p(K_{\varepsilon_1}) \xrightarrow{f_1} \cdots \xrightarrow{f_{m-1}} H_p(K_{\varepsilon_m}),
\end{equation}
where each \( H_p(K_{\varepsilon_i}) \) denotes the \( p \)-th homology group of \( K_{\varepsilon_i} \), and each \( f_i \) represents the group homomorphism induced by the inclusion map \( K_{\varepsilon_i} \subseteq K_{\varepsilon_{i+1}} \)~\cite{cal_betti}. For each \( \varepsilon_i \), the homology \( H_p(K_{\varepsilon_i}) \) (where \( p = 0, 1, \ldots \)) captures global features of the simplicial complex \( K_{\varepsilon_i} \). For example, as an abelian group, each generator of the \( 0 \)-th homology of \( K_{\varepsilon_i} \) corresponds to a (path-)connected component in \( K_{\varepsilon_i} \), each generator of the \( 1 \)-st homology corresponds to a loop structure enclosed by \( 1 \)-simplices in \( K_{\varepsilon_i} \), and each generator of the \( 2 \)-nd homology corresponds to a two-dimensional void enclosed by \( 2 \)-simplices in \( K_{\varepsilon_i} \), and so on. In particular, for a simplicial complex \( K \) consisting of finitely many simplices, the rank of the \( p \)-th homology group \( H_p(K) \)~\cite{dummit2003abstract}, known as the \( p \)-th Betti number and denoted by \( \beta_p(K) \), or simply \( \beta_p \) when \( K \) is specified, counts the number of \( p \)-dimensional "holes." Specifically, \( \beta_0 \) corresponds to the number of (path-)connected components, while \( \beta_1 \) represents the number of loops or one-dimensional holes in the complex~\cite{cal_betti,greenberg2019algebraic}. As the filtration level increases, the algebraic objects $H_p(K_{\varepsilon_i})$ track the changes in topological structures throughout the continuously varying filtration.
 
\paragraph{PH feature extraction}
The third step involves extracting topological features by summarizing the PH derived from the proposed DTFE-based filtration. In the context of TDA, these features capture properties of the dataset based on its shape and connectivity, rather than specific coordinates or values. Various methods exist for summarizing PH, including persistence barcodes and diagrams~\cite{PH_features1,PH_features2,PH_features3}, persistence landscapes~\cite{PH_features4}, persistence images~\cite{PH_features5}, and persistence curves~\cite{PH_features6}, among others. 

In this work, we utilize the Betti curves, a common summary of persistence barcodes and diagrams, serving as a specialized version of persistence curves, to extract topological information from the DTFE-based filtration. Specifically, for each $\varepsilon_i$ in the filtration described in Eq. \eqref{Eq. built filtration of simplicial complexes} and the corresponding PH in Eq. \eqref{Eq. PH of filtration}, we compute the $0$-th and $1$-st Betti numbers: $\beta_0 = {\rm rank \ } H_0(K_{\varepsilon_i})$ and $\beta_1 = {\rm rank \ } H_1(K_{\varepsilon_i})$. These Betti numbers are recorded as a topological summary of the simplicial complex at each $\varepsilon_i$ within the PH. In our analysis, we focus on the topological state of each DTFE-based simplex at the moment indexed by the filtration levels $\varepsilon$, and featurize it using homology. Leveraging other PH features, such as persistence landscapes, persistence images, and persistence curves, will be our future research directions.

\paragraph{Mathematical details}
We schematically elaborate on the essential foundation of the homology theory of simplicial complexes, which forms the basis for persistent homology. More detailed mathematical settings and theories can be found in~\cite{cal_betti}. 

Given a simplicial complex $K$ within $\mathbb{R}^n$ and a non-negative integer $p$, the $p$-th homology group of $K$, denoted as $H_p(K)$, is defined through the following steps. First, every \( p \)-simplex with \( p \geq 1 \) is equipped with two orientations. Namely, two ordered sequences \( v_0, v_1, \dots, v_p \) and \( w_0, w_1, \dots, w_p \) of vertices in \( K \) are said to be equivalent if \( w_i = v_{\phi(i)} \) for some even permutation \( \phi: \{ 0, 1, \dots, p \} \rightarrow \{ 0, 1, \dots, p \} \). For example, \( v_0, v_1, v_2 \) and \( v_1, v_2, v_0 \) have the same orientation, since the following swapping procedure can generate the latter: \( v_0, v_1, v_2 \rightarrow v_1, v_0, v_2 \rightarrow v_1, v_2, v_0 \). On the other hand, in this example, the sequences \( v_0, v_1, v_2 \) and \( v_1, v_0, v_2 \) have different orientations, as a single swap of the first one obtains the latter. In particular, an oriented $p$-simplex is represented as the equivalence class $[v_0, v_1, ..., v_n]$ of sequences of vertices in $K$. An oriented simplex \( \tau \) is defined as the additive inverse of another simplex \( \sigma \), denoted as \( \tau = -\sigma \), if \( \sigma \) and \( \tau \) have the same vertices but opposite orientations.

Second, the elementary \( p \)-chain \( c \) corresponding to an oriented \( p \)-simplex \( \sigma \) is a function $c$ from the set of all oriented \( p \)-simplices to \( \mathbb{Z} \) defined by
\begin{equation}
c(\tau) = \begin{cases}
1 &, \ \tau = \sigma \\
-1 &, \ \tau = -\sigma \\
0 &, \ \text{otherwise}.
\end{cases}
\end{equation}
Then, the $p$-th chain group of simplicial complex $K$, denoted as $C_p(K)$, is defined as the free abelian group generated by all elementary $p$-chains. Furthermore, group $C_0(K)$ is defined as the free abelian group generated by vertices, and $C_{-1}(K)$ is defined as the trivial group. 

Finally, the \( p \)-th homology of the simplicial complex \( K \) is defined by considering the boundary relations of simplices in \( K \). Specifically, for each \( p \geq 1 \), the \( p \)-th boundary map \( \partial_p: C_p(K) \rightarrow C_{p-1}(K) \) is defined as the group homomorphism such that
\begin{equation}
\partial_{p} \sigma=\partial_{p}\left[v_{0}, \ldots v_{p}\right]=\sum_{i=0}^{p}(-1)^{i}\left[v_{0}, \ldots, \widehat{v_{i}}, \ldots v_{p}\right]
\end{equation}
for every oriented $p$-simplicies. In particular, one sees that \( \partial_{p-1} \circ \partial_p = 0 \) for every \( p \geq 1 \). Consequently, the \( p \)-th homology group is defined as the quotient group \( H_p(K) = Z_p(K)/B_p(K) \), where \( Z_p(K) \) is the kernel of \( \partial_p : C_p(K) \to C_{p-1}(K) \), known as the group of \( p \)-cycles, and \( B_p(K) \) is the image of \( \partial_{p+1} : C_{p+1}(K) \to C_p(K) \), known as the group of \( p \)-boundaries. The \( p \)-th Betti number is the rank of the \( p \)-th homology group of that collection.

\paragraph{TopoPointNet architecture}
By integrating the persistent homology of TDA into a deep learning method, we construct a TopoPointNet, as illustrated in Fig.~\ref{Fig:TopoNet}, to identify and extract weak intermittency signals from the background. This network consists of two main modules. The first TDA module is designed to extract topological features from the input point cloud data of the 2D momentum for final-state particles. The second ClassifyPointNet module builds a point cloud neural network to learn the spatial encoding of these topological features, enabling the classification of signal and background events. The topological features extracted in the first module serve as inputs for the subsequent ClassifyPointNet analysis.

In the detailed architecture of TopoPointNet shown in Fig.~\ref{Fig:TopoNet}, we first construct simplicial complexes using persistent homology to capture the relationship between Betti numbers of different dimensions by varying filtration levels. In the first module, the process begins with performing a Delaunay triangulation using the scipy.spatial.Delaunay function from the advanced computational library SciPy. This is followed by extracting the data points associated with each triangle in the triangulation. The next step involves applying PH through the Geometry Understanding in Higher Dimensions (GUDHI) library, version 3.10.1. Specifically, the filtration of the simplicial complex obtained from the triangulation is carried out at different filtration levels, and the corresponding topological features are computed for each level. This is mainly done by constructing the simplicial complex with the gudhi.SimplexTree function and then applying PH using the gudhi.SimplexTree.persistence function. Subsequently, we design a two-layer one-dimensional convolutional network to extract local features, with 128 and 256 convolutional kernels, each having a kernel size of 5. This is followed by global max pooling to integrate global features and capture key information from the feature map. Then, we combine the features extracted by the previous network through two fully connected layers, providing a large number of learnable parameters to enable the network to adapt to the training data and make accurate predictions. During the fully connected process, we use the Dropout algorithm, which randomly drops a certain proportion of neurons during training to mitigate complex dependencies among neurons and prevent overfitting. The dropout rate during training is set to 0.3. Each layer applies ReLU and batch normalization, with ReLU enhancing the nonlinear mapping capability by adding an activation function, and batch normalization standardizing the input data of each batch to accelerate model training and improve generalization. Finally, the output layer utilizes the softmax function for classification, which outputs the probabilities of the input event being either a signal or a background event. 

The TopoPointNet is trained using a supervised learning technique. The training process involves feeding labeled data into the network and calculating the predicted values via forward propagation. The difference between the predicted values and the actual labels is quantified using the CrossEntropy loss function. The weights and biases of the neurons are updated using the gradient descent algorithm. By iteratively repeating this process, the difference between the predicted values and the actual labels is gradually minimized, thus improving the ability of the model to accurately predict the categories of input data.

%%%%%%%%%%%%%%%%%%%%%%%%%%%%%%%%%%%%%%%%%%%%%%%%%%%%%%%%%%%%%%%%%%%%%%%%%%%%%%%
\section{Identifying intermittency signals by extracting topological features}\label{sec:result}
%%%%%%%%%%%%%%%%%%%%%%%%%%%%%%%%%%%%%%%%%%%%%%%%%%%%%%%%%%%%%%%%%%%%%%%%%%%%%%%
\begin{figure}
    \centering     
    \includegraphics[width=1.0\linewidth]{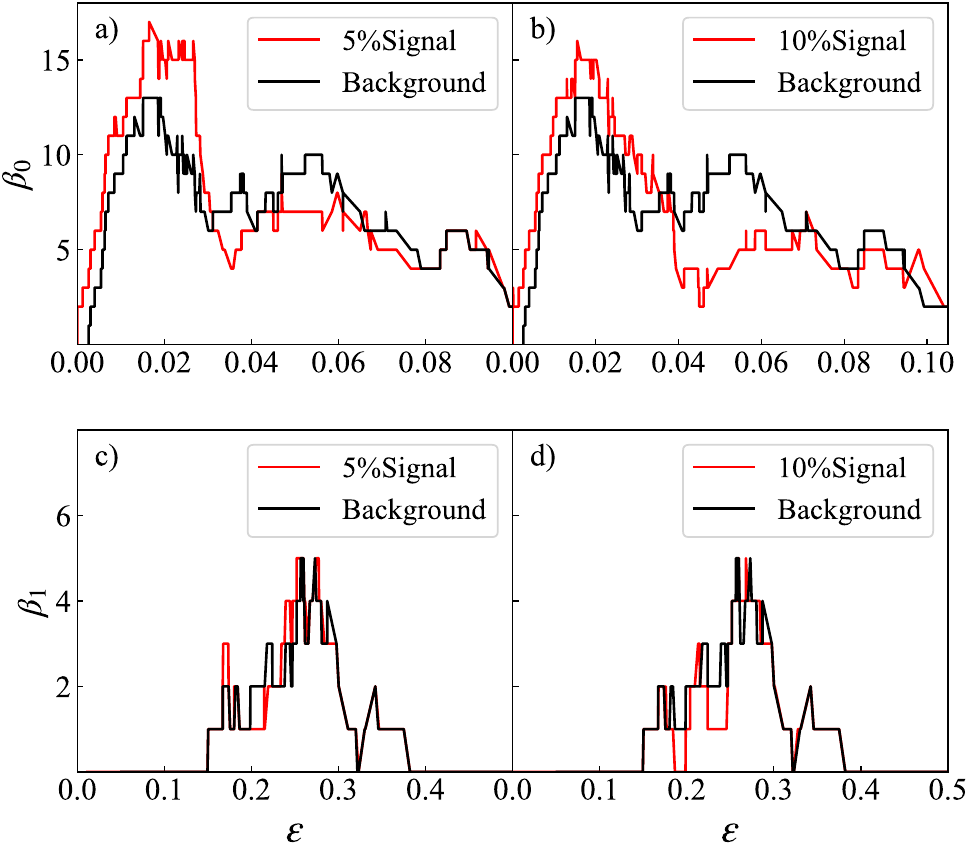}
    \caption{The distribution of Betti numbers as a function of the filtration level. The black curves represent results from the background event, while the red curves correspond to the signal event with 5\% (left panel) and 10\% (right panel) replacement ratios.}
    \label{Fig:single}
    \end{figure}
\begin{figure}
    \centering     
    \includegraphics[width=1.0\linewidth]{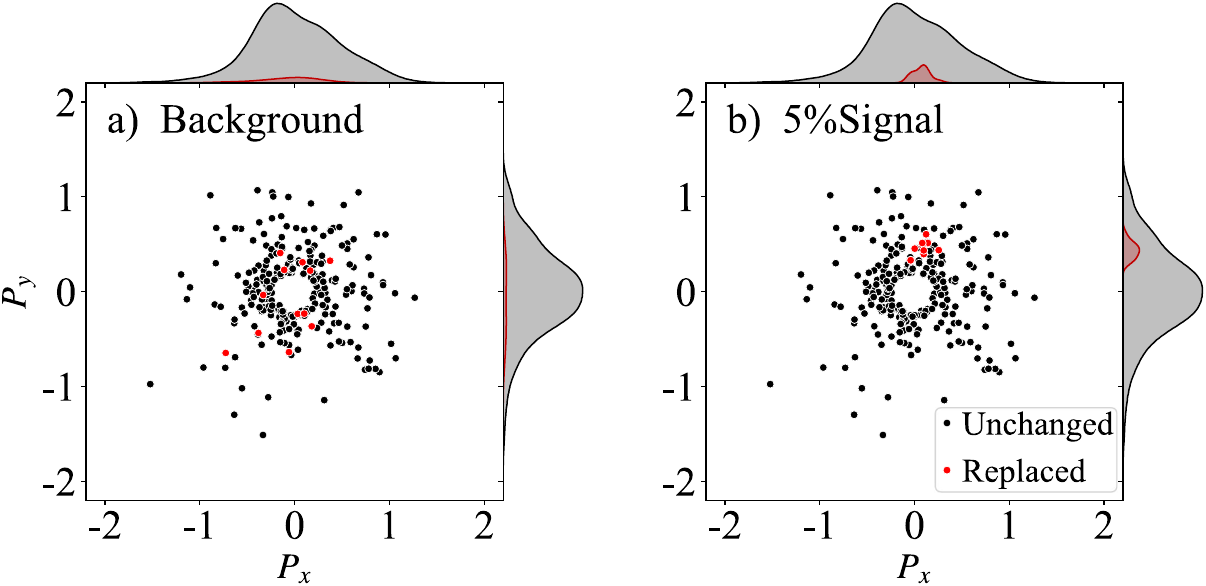}
    \caption{The two-dimensional momentum distributions from (a) background and (b) 5\% signal events, respectively. The black dots represent the unchanged UrQMD particles, while the red dots indicate the replaced critical particles introduced by the CMC model with a replacement ratio of $\lambda = 5\%$.}
    \label{Fig:replace}
    \end{figure}

In intermittency analysis, background events are common due to the inherent complexity and multi-scale nature of the system under investigation~\cite{intermittency_and_fractal_dimension3}. The trivial fluctuations and correlations present in these events may arise from various factors, such as large statistical fluctuations, external disturbances, long-range correlations, or heavy-tailed distributions. They can distort scaled factorial moments, alter scaling behaviors, and even mask or modify the true intermittent features of the data. Identifying and effectively filtering out these background events is essential for isolating the signal events that reflect the underlying intermittent dynamics of interest. This task becomes particularly challenging when the critical fluctuations in signal events are weak, as seen in experiments like NA49, NA61, and STAR.  To address this challenge, we employ the topological machine learning approach, as detailed in Sec.~\ref{sec:net}. Our method starts by extracting topological features from point cloud data of both signal and background events, using persistent homology from TDA. Since the input data are the two-dimensional momenta of each particle, the highest dimension of the topological complex formed is also two-dimensional. Therefore, we focus on the topological features of Betti numbers $\beta_0$ and $\beta_1$.

\begin{figure*}
    \centering 
    \includegraphics[scale=0.65]{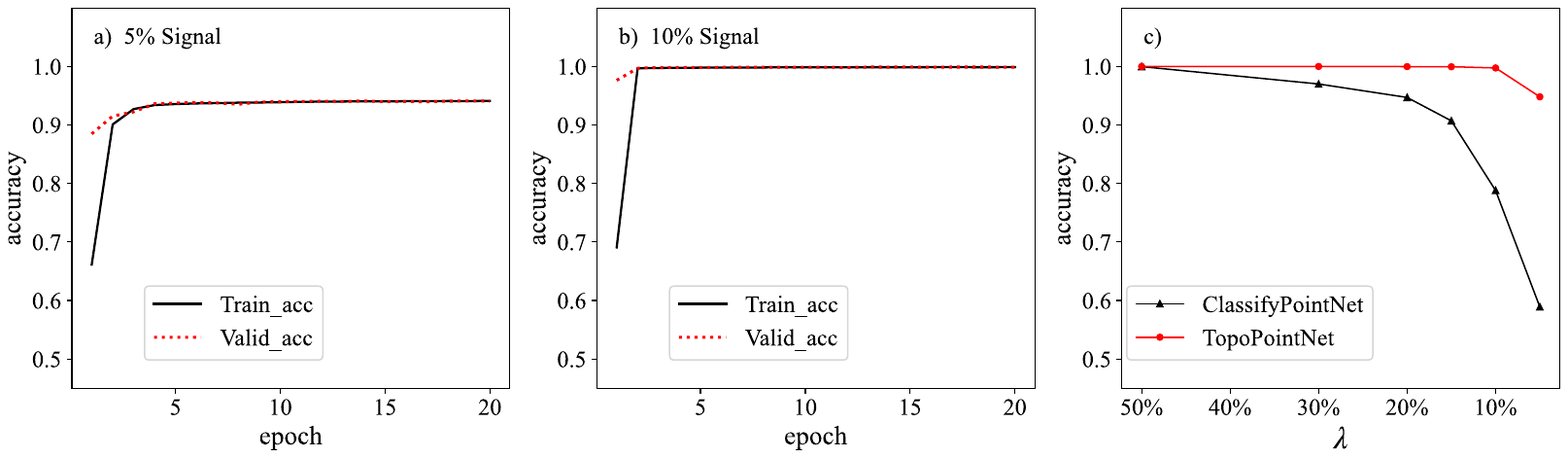}
    \caption{The training and validation accuracy as a function of epochs with (a) $\lambda = 5\%$ and (b) $\lambda = 10\%$, respectively. (c) presents a comparison of test set accuracy between pure ClassifyPointNet and TopoPointNet with various replacement ratios.}
    \label{Fig:accuracy}
    \end{figure*}
    
We begin our analysis by exploring how the Betti numbers of the signal event differ from those of the background event. In the upper panels of Fig.~\ref{Fig:single}, we show the distributions of the 0th topological features $\beta_0$ with respect to the PH filtration level $\varepsilon$. The black curves represent results from background events, while the red ones correspond to signal events with 5\% and 10\% replacement ratios in (a) and (b), respectively. We observe that both black and red curves increase with increasing $\varepsilon$ when the filtration level is small. In this region, the $\beta_0$ values from the signal event are clearly higher than those from the background event, with the red lines showing a higher peak compared to the black ones. Conversely, both the red and black curves decrease when $\varepsilon$ is large, and the $\beta_0$ values from the signal event are smaller than those from the background event. The lower panels of Fig.~\ref{Fig:single} (c) and (d) present the results for the 1st topological features in two different event samples. When $\varepsilon$ is less than 0.15, the values of $\beta_1$ are consistently 0 for both the background and signal events. In the region where $\varepsilon > 0.15$, it becomes difficult to distinguish distinct characteristics of $\beta_1$ between the background event (black curve) and the signal event (red curve). No significant differences are observed between the two types of events, with either 5\% or 10\% replacement ratios.

To understand the observed behaviors of Betti numbers, in Fig.~\ref{Fig:replace} we illustrate the momentum distribution of both the background and signal events. In Fig.~\ref{Fig:replace} (a), all data points from a background event generated by the UrQMD model are somehow evenly scattered within the chosen momentum window. The red points indicate the randomly selected 5\% of particles that will be replaced by CMC particles. The momentum distribution for the corresponding signal event, after replacing the chosen 5\% of particles as detailed in Sec. \ref{sec:event}, is depicted in Fig.~\ref{Fig:replace} (b). Here, the replaced particles (red points) gather around, exhibiting a formation similar to a cluster. Following the DTFE process described in Fig.~\ref{Fig:filtration}, we construct topological simplicial complexes on different event samples. At the early stage of filtration, the DTFE begins by connecting data points that are born as the filtering level increases, resulting in the formation of multiple complexes. The number of connected components, i.e., $\beta_0$, starts at 0 and increases with increasing $\varepsilon$. In this early stage, the signal event will emerge with more data points compared to the background event since the replaced particles (shown as red dots in Fig.~\ref{Fig:replace}) cluster together in the signal event sample. This leads to a larger $\beta_0$ in the signal event than in the background event. As the filtration level continues to increase, connections may form between clusters, causing $\beta_0$ to decrease in the late stage. During this stage, $\beta_0$ is smaller in the signal event because the number of components in the signal event decreases more rapidly compared to the background event. Eventually, all data points of the entire event form a single complete component when $\varepsilon$ becomes large enough. As for the 1st Betti number $\beta_1$, it counts the number of 1-dimensional holes or loops in a space. The observed vanishing of $\beta_1$ at $\varepsilon < 0.15$ for both background and signal events in Fig.~\ref{Fig:single} (c) and (d) indicates that holes cannot form in this region. During the PH filtering process, as the filtration level $\varepsilon$ increases, the newly born data points quickly connect with each other and soon fill into small triangular complexes because they are very close to each other. There are not enough unfilled data points available to form a topological hole in this case. As the filtration level continues to increase and surpasses a certain threshold, more data points and filled triangular complexes emerge, making it possible for holes to form. However, there are no significant differences in $\beta_1$ between background and signal events because the proportion of signal particles in this stage is too small to create a noticeable effect.

The observed differences in the 0th Betti number between the two types of events give us confidence in using this topological feature to identify weak signal events from background events in heavy-ion collisions. We select 100 arrays consisting of various filtering levels that induce changes in the spatial structure and the corresponding $\beta_0$ as input data for the point cloud network in the second module of the TopoPointNet as introduced in Fig.~\ref{Fig:TopoNet}. 

The training (black solid line) and validation (red dashed line) accuracy as a function of training epochs for 5\% and 10\% signal events are shown in Fig.~\ref{Fig:accuracy} (a) and (b), respectively. We observe that both the black and red lines initially increase with more iteration epochs and then quickly stabilize at a certain value. The validation accuracy for the 5\% signal events reaches a maximum of 94.81\%. For the 10\% signal events, the accuracy stabilizes faster than that for the 5\% signal events, with a maximum value of 99.88\%. The testing accuracy is 94.69\% for the 5\% replacement ratio and 99.85\% for the 10\% replacement ratio, both of which are very close to the corresponding validation accuracy. These results confirm that topological machine learning can classify both types of events with high accuracy.

The difference in the 2D momentum distribution between background and signal events is highly dependent on the choice of replacement ratio. In heavy-ion collisions, identifying signal events from background events becomes challenging if the replacement ratio is too small. The red curve in Fig.~\ref{Fig:accuracy} (c) illustrates the testing accuracy of TopoPointNet in detecting signal events at different replacement ratios. For comparison, we also train a pure point cloud network without the TDA module, shown as the black curve in the same figure. In this scenario, the 2D momentum of each particle is directly fed into the 1D CNN in the second module of the network architecture, as depicted in Fig.~\ref{Fig:TopoNet}, for classification. The red line remains almost unchanged with high testing accuracy when $\lambda > 10\%$. After that, it starts to gradually decrease with even smaller replacement ratios. Whereas the testing accuracy of the ClassifyPointNet without TDA decreases significantly faster with decreasing $\lambda$ compared to TopoPointNet, reaching 58.89\% for 5\% signal events. To further strengthen our conclusions, we have included two widely recognized point cloud networks: PointNet~\cite{PointNet} and PointNN~\cite{PointNN}. We adopted the same network architecture as described in the references. The results for 5\% signal events indicate that the testing accuracies were 60.91\% for PointNet and 54.87\% for PointNN. This suggests that both PointNet and PointNN exhibit lower performance compared to TopoPointNet. Therefore, topological machine learning demonstrates a markedly superior capability to detect weak signals of intermittency fluctuations compared to pure point cloud neural networks.

\begin{figure}
    \centering 
    \includegraphics[width=1\linewidth]{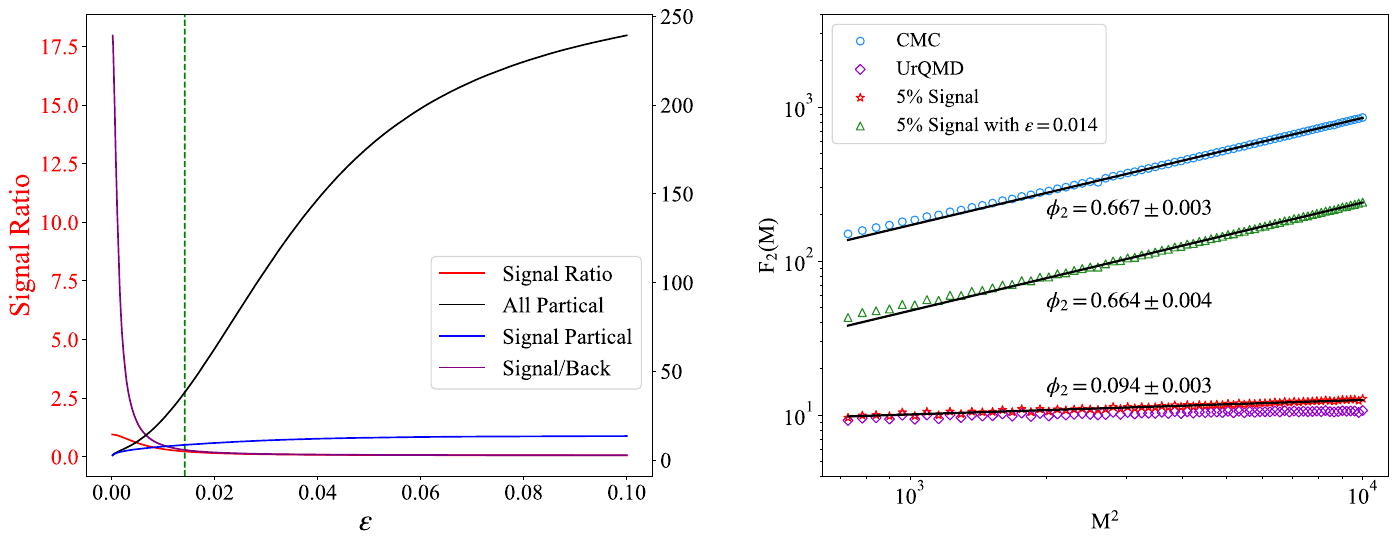}
    \caption{The second-order scaled factorial moments as a function of the number of divided cells for data samples from the pure CMC model (blue circles), UrQMD model (purple diamonds), 5\% signal events (red stars), and 5\% signal events with $\varepsilon = 0.014$  (green triangles), respectively. The statistical uncertainties, calculated by the bootstrap method, are found to be smaller than the size of the data points.}
    \label{Fig:intermittency}
    \end{figure}

Finally, we perform the intermittency analysis, as outlined in Sec. \ref{sec:event}, on different event samples. Fig.~\ref{Fig:intermittency} shows the dependence of the second-order SFMs on $M^2$, calculated for the pure CMC model (blue circles), UrQMD model (purple diamonds), 5\% signal events (red stars), and 5\% signal events with $\varepsilon = 0.014$  (green triangles). The black solid lines represent the power-law fitting according to Eq.~\eqref{Eq:PowerLaw}. It can be seen from the figure that the blue circles exhibit a strong power-law behavior as the number of partitioned cells increases. The slope of the fit, corresponding to the second-order intermittency index $\phi_2$, is determined to be $0.667\pm0.003$, which is consistent with the theoretical expectation of $\phi_2=2/3$~\cite{intermittencyindex}. This result confirms that the CMC model effectively reproduces the self-similar intermittency behavior. The purple diamonds representing results from the UrQMD background data sample show a flat trend as $M^2$ increases, which is expected since this transport model does not incorporate any critical-related intermittency mechanisms~\cite{CMC4,CMC5,critical_fluctuation1}. The red stars exhibit a gradual rise with increasing $M^2$, with the fitted intermittency index $\phi_2 = 0.094\pm0.003$, which is slightly above zero but significantly lower than the result from the pure CMC model, from which the 5\% signal particles originate. This is because, in this 5\% signal event sample,  95\% of the particles come from the UrQMD model. The dominant background particles obscure and distort the contribution of signal particles, leading to a significant underestimation of the intermittency index when calculated directly from weak signal events. This issue is also the challenge currently encountered by the NA49, NA61, or STAR collaborations in experimentally measuring weak signals of intermittency in heavy-ion collisions~\cite{CMC4}.

In the topological machine learning analysis, the filtration level is a key variable that controls the process of constructing a sequence of simplicial complexes. This allows for the analysis of data across multiple scales and uncovers important topological features of the dataset. During the Delaunay triangulation filtration process for the 5\% signal event, most particles that appear in the early stage are signal particles, as they are densely clustered in phase space. As the filtration level increases, the particles that emerge in the late stage are predominantly background particles. Therefore, by selecting an appropriate filtration level for truncation, the proportion of signal particles can be significantly increased since a large number of background particles that emerge at large $\varepsilon$ are effectively discarded. The green triangles in Fig.~\ref{Fig:intermittency} illustrate the calculated SFMs for the 5\% signal event sample, with a filtration level of $\varepsilon = 0.014$. These data points follow a good power-law behavior as $F_{2}(M)$ increases with $M^{2}$. The calculated intermittency index $\phi_{2}$ is $0.664\pm0.004$, which is consistent with the value derived directly from the pure CMC model within statistical uncertainties. It infers that by integrating the topological machine learning and  topological character selection to filter signal events and particles in the traditional intermittency analysis, we can accurately extract the intermittency index even for weak signal event samples.

%%%%%%%%%%%%%%%%%%%%%%%%%%%%%%%%%%%%%%%%%%%%%%%%%%%%%%%%%%%%%%%%%%%%%%%%%%%%%%%
\section{SUMMARY AND OUTLOOK}\label{sec:summary}
%%%%%%%%%%%%%%%%%%%%%%%%%%%%%%%%%%%%%%%%%%%%%%%%%%%%%%%%%%%%%%%%%%%%%%%%%%%%%%%
In this work, we have demonstrated the application of persistent homology from TDA to analyze and extract topological features from event samples containing critical fluctuations of intermittency in heavy-ion collisions. We have observed a clear difference in the 0th Betti number between background and weak signal events. By constructing a point cloud neural network with TDA, we have achieved a testing accuracy of 94.69\% when only 5\% of UrQMD background particles in each event are replaced by CMC signal particles. Removing the TDA module will significantly reduce performance. By selecting a filtration level of $\varepsilon = 0.014$, the calculated second-order intermittency index for the 5\% signal events closely matches
that for the pure CMC signal events.

As a first attempt to identify weak intermittency signals using cutting-edge topological machine learning in heavy-ion collisions, we have utilized a supervised learning method, focusing our analysis on two-dimensional momentum space. Extending this study to three-dimensional space in future work is highly anticipated, as it would allow for the exploration of higher-dimensional topological features, such as the 2nd Betti number. Additionally, integrating other machine learning techniques, such as unsupervised learning, could broaden the applicability of this approach to complex systems. This advancement would facilitate the use of topological machine learning in current experimental investigations on critical fluctuations of intermittency, enhance our understanding of the topological characteristics of criticality, and potentially uncover new physical phenomena in relativistic heavy-ion collisions.

%%%%%%%%%%%%%%%%%%%%%%%%%%%%%%%%%%%%%%%%%%%%%%%%%%%%%%%%%%%%%%%%%%%%%%%%%%%%%%%
\section*{Acknowledgments}
We are grateful to Prof. Xin-Nian Wang, Prof. Xiaofeng Luo, Prof. Long-Gang Pang, Dr. Xue Gong, and Longlong Li for fruitful discussions and comments. The numerical simulations have been performed on the GPU cluster in the Nuclear Science Computing Center at Central China Normal University ($\rm{NSC^3}$). This work is supported by the National Key Research and Development Program of China (No. 2024YFE0110103 and No. 2022YFA1604900), the National Natural Science Foundation of China (No. 12275102) and the Central China Normal University Student Innovation and Entrepreneurship Training Program Project.
%%%%%%%%%%%%%%%%%%%%%%%%%%%%%%%%%%%%%%%%%%%%%%%%%%%%%%%%%%%%%%%%%%%%%%%%%%%%%%%
\bibliography{bib.bib}		

\begin{thebibliography}{85}%
\makeatletter
\providecommand \@ifxundefined [1]{%
 \@ifx{#1\undefined}
}%
\providecommand \@ifnum [1]{%
 \ifnum #1\expandafter \@firstoftwo
 \else \expandafter \@secondoftwo
 \fi
}%
\providecommand \@ifx [1]{%
 \ifx #1\expandafter \@firstoftwo
 \else \expandafter \@secondoftwo
 \fi
}%
\providecommand \natexlab [1]{#1}%
\providecommand \enquote  [1]{``#1''}%
\providecommand \bibnamefont  [1]{#1}%
\providecommand \bibfnamefont [1]{#1}%
\providecommand \citenamefont [1]{#1}%
\providecommand \href@noop [0]{\@secondoftwo}%
\providecommand \href [0]{\begingroup \@sanitize@url \@href}%
\providecommand \@href[1]{\@@startlink{#1}\@@href}%
\providecommand \@@href[1]{\endgroup#1\@@endlink}%
\providecommand \@sanitize@url [0]{\catcode `\\12\catcode `\$12\catcode `\&12\catcode `\#12\catcode `\^12\catcode `\_12\catcode `\%12\relax}%
\providecommand \@@startlink[1]{}%
\providecommand \@@endlink[0]{}%
\providecommand \url  [0]{\begingroup\@sanitize@url \@url }%
\providecommand \@url [1]{\endgroup\@href {#1}{\urlprefix }}%
\providecommand \urlprefix  [0]{URL }%
\providecommand \Eprint [0]{\href }%
\providecommand \doibase [0]{http://dx.doi.org/}%
\providecommand \selectlanguage [0]{\@gobble}%
\providecommand \bibinfo  [0]{\@secondoftwo}%
\providecommand \bibfield  [0]{\@secondoftwo}%
\providecommand \translation [1]{[#1]}%
\providecommand \BibitemOpen [0]{}%
\providecommand \bibitemStop [0]{}%
\providecommand \bibitemNoStop [0]{.\EOS\space}%
\providecommand \EOS [0]{\spacefactor3000\relax}%
\providecommand \BibitemShut  [1]{\csname bibitem#1\endcsname}%
\let\auto@bib@innerbib\@empty
%</preamble>
\bibitem [{\citenamefont {Stephanov}\ \emph {et~al.}(1998)\citenamefont {Stephanov}, \citenamefont {Rajagopal},\ and\ \citenamefont {Shuryak}}]{StephanovPD}%
  \BibitemOpen
  \bibfield  {author} {\bibinfo {author} {\bibfnamefont {M.}~\bibnamefont {Stephanov}}, \bibinfo {author} {\bibfnamefont {K.}~\bibnamefont {Rajagopal}}, \ and\ \bibinfo {author} {\bibfnamefont {E.~V.}\ \bibnamefont {Shuryak}},\ }\href {\doibase 10.1103/PhysRevLett.81.4816} {\bibfield  {journal} {\bibinfo  {journal} {Phys. Rev. Lett.}\ }\textbf {\bibinfo {volume} {81}},\ \bibinfo {pages} {4816} (\bibinfo {year} {1998})}\BibitemShut {NoStop}%
\bibitem [{\citenamefont {Adams}\ \emph {et~al.}(2005)\citenamefont {Adams} \emph {et~al.}}]{adams2005experimental}%
  \BibitemOpen
  \bibfield  {author} {\bibinfo {author} {\bibfnamefont {J.}~\bibnamefont {Adams}} \emph {et~al.} (\bibinfo {collaboration} {STAR Coll.}),\ }\href {\doibase 10.1016/j.nuclphysa.2005.03.085} {\bibfield  {journal} {\bibinfo  {journal} {Nucl. Phys. A}\ }\textbf {\bibinfo {volume} {757}},\ \bibinfo {pages} {102} (\bibinfo {year} {2005})}\BibitemShut {NoStop}%
\bibitem [{\citenamefont {Asakawa}\ \emph {et~al.}(2000)\citenamefont {Asakawa}, \citenamefont {Heinz},\ and\ \citenamefont {M{\"u}ller}}]{conservecharge0}%
  \BibitemOpen
  \bibfield  {author} {\bibinfo {author} {\bibfnamefont {M.}~\bibnamefont {Asakawa}}, \bibinfo {author} {\bibfnamefont {U.}~\bibnamefont {Heinz}}, \ and\ \bibinfo {author} {\bibfnamefont {B.}~\bibnamefont {M{\"u}ller}},\ }\href {\doibase 10.1103/PhysRevLett.85.2072} {\bibfield  {journal} {\bibinfo  {journal} {Phys. Rev. Lett.}\ }\textbf {\bibinfo {volume} {85}},\ \bibinfo {pages} {2072} (\bibinfo {year} {2000})}\BibitemShut {NoStop}%
\bibitem [{\citenamefont {Koch}\ \emph {et~al.}(2005)\citenamefont {Koch}, \citenamefont {Majumder},\ and\ \citenamefont {Randrup}}]{conservecharge1}%
  \BibitemOpen
  \bibfield  {author} {\bibinfo {author} {\bibfnamefont {V.}~\bibnamefont {Koch}}, \bibinfo {author} {\bibfnamefont {A.}~\bibnamefont {Majumder}}, \ and\ \bibinfo {author} {\bibfnamefont {J.}~\bibnamefont {Randrup}},\ }\href {\doibase 10.1103/PhysRevLett.95.182301} {\bibfield  {journal} {\bibinfo  {journal} {Phys. Rev. Lett.}\ }\textbf {\bibinfo {volume} {95}},\ \bibinfo {pages} {182301} (\bibinfo {year} {2005})}\BibitemShut {NoStop}%
\bibitem [{\citenamefont {Aoki}\ \emph {et~al.}(2006)\citenamefont {Aoki}, \citenamefont {Endr{\H{o}}di}, \citenamefont {Fodor} \emph {et~al.}}]{Lattice1}%
  \BibitemOpen
  \bibfield  {author} {\bibinfo {author} {\bibfnamefont {Y.}~\bibnamefont {Aoki}}, \bibinfo {author} {\bibfnamefont {G.}~\bibnamefont {Endr{\H{o}}di}}, \bibinfo {author} {\bibfnamefont {Z.}~\bibnamefont {Fodor}},  \emph {et~al.},\ }\href {\doibase 10.1038/nature05120} {\bibfield  {journal} {\bibinfo  {journal} {Nature}\ }\textbf {\bibinfo {volume} {443}},\ \bibinfo {pages} {675} (\bibinfo {year} {2006})}\BibitemShut {NoStop}%
\bibitem [{\citenamefont {Bhattacharya}\ \emph {et~al.}(2014)\citenamefont {Bhattacharya}, \citenamefont {Buchoff}, \citenamefont {Christ} \emph {et~al.}}]{Lattice2}%
  \BibitemOpen
  \bibfield  {author} {\bibinfo {author} {\bibfnamefont {T.}~\bibnamefont {Bhattacharya}}, \bibinfo {author} {\bibfnamefont {M.~I.}\ \bibnamefont {Buchoff}}, \bibinfo {author} {\bibfnamefont {N.~H.}\ \bibnamefont {Christ}},  \emph {et~al.},\ }\href {\doibase 10.1103/PhysRevLett.113.082001} {\bibfield  {journal} {\bibinfo  {journal} {Phys. Rev. Lett.}\ }\textbf {\bibinfo {volume} {113}},\ \bibinfo {pages} {082001} (\bibinfo {year} {2014})}\BibitemShut {NoStop}%
\bibitem [{\citenamefont {Stephanov}(2005)}]{CEP1}%
  \BibitemOpen
  \bibfield  {author} {\bibinfo {author} {\bibfnamefont {M.}~\bibnamefont {Stephanov}},\ }\href {\doibase 10.1142/S0217751X05027965} {\bibfield  {journal} {\bibinfo  {journal} {Int. J. Mod. Phys. A}\ }\textbf {\bibinfo {volume} {20}},\ \bibinfo {pages} {4387} (\bibinfo {year} {2005})}\BibitemShut {NoStop}%
\bibitem [{\citenamefont {Fodor}\ and\ \citenamefont {Katz}(2004)}]{CEP2}%
  \BibitemOpen
  \bibfield  {author} {\bibinfo {author} {\bibfnamefont {Z.}~\bibnamefont {Fodor}}\ and\ \bibinfo {author} {\bibfnamefont {S.~D.}\ \bibnamefont {Katz}},\ }\href {\doibase 10.1088/1126-6708/2004/04/050} {\bibfield  {journal} {\bibinfo  {journal} {JHEP}\ }\textbf {\bibinfo {volume} {2004}},\ \bibinfo {pages} {050} (\bibinfo {year} {2004})}\BibitemShut {NoStop}%
\bibitem [{\citenamefont {Gavai}\ and\ \citenamefont {Gupta}(2005)}]{CEP3}%
  \BibitemOpen
  \bibfield  {author} {\bibinfo {author} {\bibfnamefont {R.~V.}\ \bibnamefont {Gavai}}\ and\ \bibinfo {author} {\bibfnamefont {S.}~\bibnamefont {Gupta}},\ }\href {\doibase 10.1103/PhysRevD.71.114014} {\bibfield  {journal} {\bibinfo  {journal} {Phys. Rev. D}\ }\textbf {\bibinfo {volume} {71}},\ \bibinfo {pages} {114014} (\bibinfo {year} {2005})}\BibitemShut {NoStop}%
\bibitem [{\citenamefont {Bzdak}\ \emph {et~al.}(2020)\citenamefont {Bzdak}, \citenamefont {Esumi}, \citenamefont {Koch} \emph {et~al.}}]{RHIC1}%
  \BibitemOpen
  \bibfield  {author} {\bibinfo {author} {\bibfnamefont {A.}~\bibnamefont {Bzdak}}, \bibinfo {author} {\bibfnamefont {S.}~\bibnamefont {Esumi}}, \bibinfo {author} {\bibfnamefont {V.}~\bibnamefont {Koch}},  \emph {et~al.},\ }\href {\doibase 10.1016/j.physrep.2020.01.005} {\bibfield  {journal} {\bibinfo  {journal} {Phys. Rep.}\ }\textbf {\bibinfo {volume} {853}},\ \bibinfo {pages} {1} (\bibinfo {year} {2020})}\BibitemShut {NoStop}%
\bibitem [{\citenamefont {Luo}\ and\ \citenamefont {Xu}(2017)}]{RHIC2}%
  \BibitemOpen
  \bibfield  {author} {\bibinfo {author} {\bibfnamefont {X.}~\bibnamefont {Luo}}\ and\ \bibinfo {author} {\bibfnamefont {N.}~\bibnamefont {Xu}},\ }\href {\doibase 10.1007/s41365-017-0257-0} {\bibfield  {journal} {\bibinfo  {journal} {Nucl. Sci. Tech.}\ }\textbf {\bibinfo {volume} {28}},\ \bibinfo {pages} {1} (\bibinfo {year} {2017})}\BibitemShut {NoStop}%
\bibitem [{\citenamefont {Bower}\ and\ \citenamefont {Gavin}(2001)}]{fluctuation1}%
  \BibitemOpen
  \bibfield  {author} {\bibinfo {author} {\bibfnamefont {D.}~\bibnamefont {Bower}}\ and\ \bibinfo {author} {\bibfnamefont {S.}~\bibnamefont {Gavin}},\ }\href {\doibase 10.1103/PhysRevC.64.051902} {\bibfield  {journal} {\bibinfo  {journal} {Phys. Rev. C}\ }\textbf {\bibinfo {volume} {64}},\ \bibinfo {pages} {051902} (\bibinfo {year} {2001})}\BibitemShut {NoStop}%
\bibitem [{\citenamefont {Antoniou}(2001)}]{fluctuation2}%
  \BibitemOpen
  \bibfield  {author} {\bibinfo {author} {\bibfnamefont {N.~G.}\ \bibnamefont {Antoniou}},\ }\href {\doibase 10.1016/S0920-5632(00)01016-1} {\bibfield  {journal} {\bibinfo  {journal} {Nucl. Phys. B, Proc. Suppl.}\ }\textbf {\bibinfo {volume} {92}},\ \bibinfo {pages} {26} (\bibinfo {year} {2001})}\BibitemShut {NoStop}%
\bibitem [{\citenamefont {Sun}\ \emph {et~al.}(2017)\citenamefont {Sun}, \citenamefont {Chen}, \citenamefont {Ko} \emph {et~al.}}]{fluctuation4}%
  \BibitemOpen
  \bibfield  {author} {\bibinfo {author} {\bibfnamefont {K.~J.}\ \bibnamefont {Sun}}, \bibinfo {author} {\bibfnamefont {L.~W.}\ \bibnamefont {Chen}}, \bibinfo {author} {\bibfnamefont {C.~M.}\ \bibnamefont {Ko}},  \emph {et~al.},\ }\href {\doibase 10.1016/j.physletb.2017.09.056} {\bibfield  {journal} {\bibinfo  {journal} {Phys. Lett. B}\ }\textbf {\bibinfo {volume} {774}},\ \bibinfo {pages} {103} (\bibinfo {year} {2017})}\BibitemShut {NoStop}%
\bibitem [{\citenamefont {Sun}\ \emph {et~al.}(2018)\citenamefont {Sun}, \citenamefont {Chen}, \citenamefont {Ko} \emph {et~al.}}]{fluctuation5}%
  \BibitemOpen
  \bibfield  {author} {\bibinfo {author} {\bibfnamefont {K.~J.}\ \bibnamefont {Sun}}, \bibinfo {author} {\bibfnamefont {L.~W.}\ \bibnamefont {Chen}}, \bibinfo {author} {\bibfnamefont {C.~M.}\ \bibnamefont {Ko}},  \emph {et~al.},\ }\href {\doibase 10.1016/j.physletb.2018.04.035} {\bibfield  {journal} {\bibinfo  {journal} {Phys. Lett. B}\ }\textbf {\bibinfo {volume} {781}},\ \bibinfo {pages} {499} (\bibinfo {year} {2018})}\BibitemShut {NoStop}%
\bibitem [{\citenamefont {Berdnikov}\ and\ \citenamefont {Rajagopal}(2000)}]{baryon_density_fluctuations2}%
  \BibitemOpen
  \bibfield  {author} {\bibinfo {author} {\bibfnamefont {B.}~\bibnamefont {Berdnikov}}\ and\ \bibinfo {author} {\bibfnamefont {K.}~\bibnamefont {Rajagopal}},\ }\href {\doibase 10.1103/PhysRevD.61.105017} {\bibfield  {journal} {\bibinfo  {journal} {Phys. Rev. D}\ }\textbf {\bibinfo {volume} {61}},\ \bibinfo {pages} {105017} (\bibinfo {year} {2000})}\BibitemShut {NoStop}%
\bibitem [{\citenamefont {Bialas}\ and\ \citenamefont {Hwa}(1991)}]{intermittency}%
  \BibitemOpen
  \bibfield  {author} {\bibinfo {author} {\bibfnamefont {A.}~\bibnamefont {Bialas}}\ and\ \bibinfo {author} {\bibfnamefont {R.~C.}\ \bibnamefont {Hwa}},\ }\href {\doibase 10.1016/0370-2693(91)91747-J} {\bibfield  {journal} {\bibinfo  {journal} {Phys. Lett. B}\ }\textbf {\bibinfo {volume} {253}},\ \bibinfo {pages} {436} (\bibinfo {year} {1991})}\BibitemShut {NoStop}%
\bibitem [{\citenamefont {Antoniou}\ \emph {et~al.}(2006)\citenamefont {Antoniou}, \citenamefont {Diakonos}, \citenamefont {Kapoyannis} \emph {et~al.}}]{CMC1}%
  \BibitemOpen
  \bibfield  {author} {\bibinfo {author} {\bibfnamefont {N.~G.}\ \bibnamefont {Antoniou}}, \bibinfo {author} {\bibfnamefont {F.~K.}\ \bibnamefont {Diakonos}}, \bibinfo {author} {\bibfnamefont {A.~S.}\ \bibnamefont {Kapoyannis}},  \emph {et~al.},\ }\href {\doibase 10.1103/PhysRevLett.97.032002} {\bibfield  {journal} {\bibinfo  {journal} {Phys. Rev. Lett.}\ }\textbf {\bibinfo {volume} {97}},\ \bibinfo {pages} {032002} (\bibinfo {year} {2006})}\BibitemShut {NoStop}%
\bibitem [{\citenamefont {Antoniou}\ \emph {et~al.}(2001)\citenamefont {Antoniou}, \citenamefont {Contoyiannis}, \citenamefont {Diakonos} \emph {et~al.}}]{CMC2}%
  \BibitemOpen
  \bibfield  {author} {\bibinfo {author} {\bibfnamefont {N.~G.}\ \bibnamefont {Antoniou}}, \bibinfo {author} {\bibfnamefont {Y.~F.}\ \bibnamefont {Contoyiannis}}, \bibinfo {author} {\bibfnamefont {F.~K.}\ \bibnamefont {Diakonos}},  \emph {et~al.},\ }\href {\doibase 10.1016/S0375-9474(01)00921-6} {\bibfield  {journal} {\bibinfo  {journal} {Nucl. Phys. A}\ }\textbf {\bibinfo {volume} {693}},\ \bibinfo {pages} {799} (\bibinfo {year} {2001})}\BibitemShut {NoStop}%
\bibitem [{\citenamefont {Antoniou}\ \emph {et~al.}(2018)\citenamefont {Antoniou}, \citenamefont {Diakonos}, \citenamefont {Maintas} \emph {et~al.}}]{CMC3}%
  \BibitemOpen
  \bibfield  {author} {\bibinfo {author} {\bibfnamefont {N.~G.}\ \bibnamefont {Antoniou}}, \bibinfo {author} {\bibfnamefont {F.~K.}\ \bibnamefont {Diakonos}}, \bibinfo {author} {\bibfnamefont {X.~N.}\ \bibnamefont {Maintas}},  \emph {et~al.},\ }\href {\doibase 10.1103/PhysRevD.97.034015} {\bibfield  {journal} {\bibinfo  {journal} {Phys. Rev. D}\ }\textbf {\bibinfo {volume} {97}},\ \bibinfo {pages} {034015} (\bibinfo {year} {2018})}\BibitemShut {NoStop}%
\bibitem [{\citenamefont {Li}(2022)}]{CMC4}%
  \BibitemOpen
  \bibfield  {author} {\bibinfo {author} {\bibfnamefont {Z.}~\bibnamefont {Li}},\ }\href {\doibase 10.1142/S0217732322300099} {\bibfield  {journal} {\bibinfo  {journal} {Mod. Phys. Lett. A}\ }\textbf {\bibinfo {volume} {37}},\ \bibinfo {pages} {2230009} (\bibinfo {year} {2022})}\BibitemShut {NoStop}%
\bibitem [{\citenamefont {Wu}\ \emph {et~al.}(2020)\citenamefont {Wu}, \citenamefont {Lin}, \citenamefont {Wu} \emph {et~al.}}]{CMC5}%
  \BibitemOpen
  \bibfield  {author} {\bibinfo {author} {\bibfnamefont {J.}~\bibnamefont {Wu}}, \bibinfo {author} {\bibfnamefont {Y.}~\bibnamefont {Lin}}, \bibinfo {author} {\bibfnamefont {Y.}~\bibnamefont {Wu}},  \emph {et~al.},\ }\href {\doibase 10.1016/j.physletb.2019.135186} {\bibfield  {journal} {\bibinfo  {journal} {Phys. Lett. B}\ }\textbf {\bibinfo {volume} {801}},\ \bibinfo {pages} {135186} (\bibinfo {year} {2020})}\BibitemShut {NoStop}%
\bibitem [{\citenamefont {Wu}\ \emph {et~al.}(2021)\citenamefont {Wu}, \citenamefont {Lin}, \citenamefont {Li} \emph {et~al.}}]{critical_fluctuation1}%
  \BibitemOpen
  \bibfield  {author} {\bibinfo {author} {\bibfnamefont {J.}~\bibnamefont {Wu}}, \bibinfo {author} {\bibfnamefont {Y.}~\bibnamefont {Lin}}, \bibinfo {author} {\bibfnamefont {Z.}~\bibnamefont {Li}},  \emph {et~al.},\ }\href {\doibase 10.1103/PhysRevC.104.034902} {\bibfield  {journal} {\bibinfo  {journal} {Phys. Rev. C}\ }\textbf {\bibinfo {volume} {104}},\ \bibinfo {pages} {034902} (\bibinfo {year} {2021})}\BibitemShut {NoStop}%
\bibitem [{\citenamefont {Wu}\ \emph {et~al.}(2022)\citenamefont {Wu}, \citenamefont {Li}, \citenamefont {Luo} \emph {et~al.}}]{critical_fluctuation2}%
  \BibitemOpen
  \bibfield  {author} {\bibinfo {author} {\bibfnamefont {J.}~\bibnamefont {Wu}}, \bibinfo {author} {\bibfnamefont {Z.}~\bibnamefont {Li}}, \bibinfo {author} {\bibfnamefont {X.}~\bibnamefont {Luo}},  \emph {et~al.},\ }\href {\doibase 10.1103/PhysRevC.106.054905} {\bibfield  {journal} {\bibinfo  {journal} {Phys. Rev. C}\ }\textbf {\bibinfo {volume} {106}},\ \bibinfo {pages} {054905} (\bibinfo {year} {2022})}\BibitemShut {NoStop}%
\bibitem [{\citenamefont {Anticic}\ \emph {et~al.}(2015)\citenamefont {Anticic} \emph {et~al.}}]{NA49}%
  \BibitemOpen
  \bibfield  {author} {\bibinfo {author} {\bibfnamefont {T.}~\bibnamefont {Anticic}} \emph {et~al.} (\bibinfo {collaboration} {NA49 Coll.}),\ }\href {\doibase 10.1140/epjc/s10052-015-3738-5} {\bibfield  {journal} {\bibinfo  {journal} {Eur. Phys. J. C}\ }\textbf {\bibinfo {volume} {75}},\ \bibinfo {pages} {1} (\bibinfo {year} {2015})}\BibitemShut {NoStop}%
\bibitem [{\citenamefont {Adhikary}\ \emph {et~al.}(2023)\citenamefont {Adhikary} \emph {et~al.}}]{NA61_150}%
  \BibitemOpen
  \bibfield  {author} {\bibinfo {author} {\bibfnamefont {H.}~\bibnamefont {Adhikary}} \emph {et~al.} (\bibinfo {collaboration} {NA61/SHINE Coll.}),\ }\href {\doibase 10.1140/epjc/s10052-023-11942-9} {\bibfield  {journal} {\bibinfo  {journal} {Eur. Phys. J. C}\ }\textbf {\bibinfo {volume} {83}},\ \bibinfo {pages} {1} (\bibinfo {year} {2023})}\BibitemShut {NoStop}%
\bibitem [{\citenamefont {Adhikary}\ \emph {et~al.}(2024)\citenamefont {Adhikary} \emph {et~al.}}]{NA61}%
  \BibitemOpen
  \bibfield  {author} {\bibinfo {author} {\bibfnamefont {H.}~\bibnamefont {Adhikary}} \emph {et~al.} (\bibinfo {collaboration} {NA61/SHINE Coll.}),\ }\href {\doibase 10.1140/epjc/s10052-024-13012-0} {\bibfield  {journal} {\bibinfo  {journal} {Eur. Phys. J. C}\ }\textbf {\bibinfo {volume} {84}},\ \bibinfo {pages} {741} (\bibinfo {year} {2024})}\BibitemShut {NoStop}%
\bibitem [{\citenamefont {Abdulhamid}\ \emph {et~al.}(2023)\citenamefont {Abdulhamid}, \citenamefont {Aboona}, \citenamefont {Adam} \emph {et~al.}}]{STARBESI}%
  \BibitemOpen
  \bibfield  {author} {\bibinfo {author} {\bibfnamefont {M.~I.}\ \bibnamefont {Abdulhamid}}, \bibinfo {author} {\bibfnamefont {B.~E.}\ \bibnamefont {Aboona}}, \bibinfo {author} {\bibfnamefont {J.}~\bibnamefont {Adam}},  \emph {et~al.},\ }\href {\doibase 10.1016/j.physletb.2023.138165} {\bibfield  {journal} {\bibinfo  {journal} {Phys. Lett. B}\ }\textbf {\bibinfo {volume} {845}},\ \bibinfo {pages} {138165} (\bibinfo {year} {2023})}\BibitemShut {NoStop}%
\bibitem [{\citenamefont {Hinton}\ and\ \citenamefont {Salakhutdinov}(2006)}]{ML1Science}%
  \BibitemOpen
  \bibfield  {author} {\bibinfo {author} {\bibfnamefont {G.~E.}\ \bibnamefont {Hinton}}\ and\ \bibinfo {author} {\bibfnamefont {R.~R.}\ \bibnamefont {Salakhutdinov}},\ }\href {\doibase 10.1126/science.1127647} {\bibfield  {journal} {\bibinfo  {journal} {Science}\ }\textbf {\bibinfo {volume} {313}},\ \bibinfo {pages} {504} (\bibinfo {year} {2006})}\BibitemShut {NoStop}%
\bibitem [{\citenamefont {LeCun}\ \emph {et~al.}(2015)\citenamefont {LeCun}, \citenamefont {Bengio},\ and\ \citenamefont {Hinton}}]{ML2Nature}%
  \BibitemOpen
  \bibfield  {author} {\bibinfo {author} {\bibfnamefont {Y.}~\bibnamefont {LeCun}}, \bibinfo {author} {\bibfnamefont {Y.}~\bibnamefont {Bengio}}, \ and\ \bibinfo {author} {\bibfnamefont {G.}~\bibnamefont {Hinton}},\ }\href {\doibase 10.1038/nature14539} {\bibfield  {journal} {\bibinfo  {journal} {Nature}\ }\textbf {\bibinfo {volume} {521}},\ \bibinfo {pages} {436} (\bibinfo {year} {2015})}\BibitemShut {NoStop}%
\bibitem [{\citenamefont {Radovic}\ \emph {et~al.}(2018)\citenamefont {Radovic}, \citenamefont {Williams}, \citenamefont {Rousseau} \emph {et~al.}}]{ML_Apply}%
  \BibitemOpen
  \bibfield  {author} {\bibinfo {author} {\bibfnamefont {A.}~\bibnamefont {Radovic}}, \bibinfo {author} {\bibfnamefont {M.}~\bibnamefont {Williams}}, \bibinfo {author} {\bibfnamefont {D.}~\bibnamefont {Rousseau}},  \emph {et~al.},\ }\href {\doibase 10.1038/s41586-018-0361-2} {\bibfield  {journal} {\bibinfo  {journal} {Nature}\ }\textbf {\bibinfo {volume} {560}},\ \bibinfo {pages} {41} (\bibinfo {year} {2018})}\BibitemShut {NoStop}%
\bibitem [{\citenamefont {Huang}\ \emph {et~al.}(2022)\citenamefont {Huang}, \citenamefont {Pang}, \citenamefont {Luo} \emph {et~al.}}]{ML_mix}%
  \BibitemOpen
  \bibfield  {author} {\bibinfo {author} {\bibfnamefont {Y.}~\bibnamefont {Huang}}, \bibinfo {author} {\bibfnamefont {L.-G.}\ \bibnamefont {Pang}}, \bibinfo {author} {\bibfnamefont {X.}~\bibnamefont {Luo}},  \emph {et~al.},\ }\href {\doibase 10.1016/j.physletb.2022.137001} {\bibfield  {journal} {\bibinfo  {journal} {Phys. Lett. B}\ }\textbf {\bibinfo {volume} {827}},\ \bibinfo {pages} {137001} (\bibinfo {year} {2022})}\BibitemShut {NoStop}%
\bibitem [{\citenamefont {Hamilton}\ \emph {et~al.}(2022)\citenamefont {Hamilton}, \citenamefont {Dore},\ and\ \citenamefont {Plumberg}}]{DTFE1}%
  \BibitemOpen
  \bibfield  {author} {\bibinfo {author} {\bibfnamefont {G.}~\bibnamefont {Hamilton}}, \bibinfo {author} {\bibfnamefont {T.}~\bibnamefont {Dore}}, \ and\ \bibinfo {author} {\bibfnamefont {C.}~\bibnamefont {Plumberg}},\ }\href {\doibase 10.1103/PhysRevC.106.064912} {\bibfield  {journal} {\bibinfo  {journal} {Phys. Rev. C}\ }\textbf {\bibinfo {volume} {106}},\ \bibinfo {pages} {064912} (\bibinfo {year} {2022})}\BibitemShut {NoStop}%
\bibitem [{\citenamefont {Steinheimer}\ \emph {et~al.}(2019)\citenamefont {Steinheimer}, \citenamefont {Pang}, \citenamefont {Zhou} \emph {et~al.}}]{pointcloud1}%
  \BibitemOpen
  \bibfield  {author} {\bibinfo {author} {\bibfnamefont {J.}~\bibnamefont {Steinheimer}}, \bibinfo {author} {\bibfnamefont {L.-G.}\ \bibnamefont {Pang}}, \bibinfo {author} {\bibfnamefont {K.}~\bibnamefont {Zhou}},  \emph {et~al.},\ }\href {\doibase 10.1007/JHEP12(2019)122} {\bibfield  {journal} {\bibinfo  {journal} {JHEP}\ }\textbf {\bibinfo {volume} {2019}},\ \bibinfo {pages} {1} (\bibinfo {year} {2019})}\BibitemShut {NoStop}%
\bibitem [{\citenamefont {Kuttan}\ \emph {et~al.}(2021)\citenamefont {Kuttan}, \citenamefont {Zhou}, \citenamefont {Steinheimer} \emph {et~al.}}]{pointcloud2}%
  \BibitemOpen
  \bibfield  {author} {\bibinfo {author} {\bibfnamefont {M.~O.}\ \bibnamefont {Kuttan}}, \bibinfo {author} {\bibfnamefont {K.}~\bibnamefont {Zhou}}, \bibinfo {author} {\bibfnamefont {J.}~\bibnamefont {Steinheimer}},  \emph {et~al.},\ }\href {\doibase 10.1007/JHEP10(2021)184} {\bibfield  {journal} {\bibinfo  {journal} {JHEP}\ }\textbf {\bibinfo {volume} {2021}},\ \bibinfo {pages} {1} (\bibinfo {year} {2021})}\BibitemShut {NoStop}%
\bibitem [{\citenamefont {Donato}\ \emph {et~al.}(2016)\citenamefont {Donato}, \citenamefont {Gori}, \citenamefont {Pettini} \emph {et~al.}}]{XY_Model}%
  \BibitemOpen
  \bibfield  {author} {\bibinfo {author} {\bibfnamefont {I.}~\bibnamefont {Donato}}, \bibinfo {author} {\bibfnamefont {M.}~\bibnamefont {Gori}}, \bibinfo {author} {\bibfnamefont {M.}~\bibnamefont {Pettini}},  \emph {et~al.},\ }\href {\doibase 10.1103/PhysRevE.93.052138} {\bibfield  {journal} {\bibinfo  {journal} {Phys. Rev. E}\ }\textbf {\bibinfo {volume} {93}},\ \bibinfo {pages} {052138} (\bibinfo {year} {2016})}\BibitemShut {NoStop}%
\bibitem [{\citenamefont {Cole}\ \emph {et~al.}(2021)\citenamefont {Cole}, \citenamefont {Loges},\ and\ \citenamefont {Shiu}}]{PHPC1}%
  \BibitemOpen
  \bibfield  {author} {\bibinfo {author} {\bibfnamefont {A.}~\bibnamefont {Cole}}, \bibinfo {author} {\bibfnamefont {G.~J.}\ \bibnamefont {Loges}}, \ and\ \bibinfo {author} {\bibfnamefont {G.}~\bibnamefont {Shiu}},\ }\href {\doibase 10.1103/PhysRevB.104.104426} {\bibfield  {journal} {\bibinfo  {journal} {Phys. Rev. B}\ }\textbf {\bibinfo {volume} {104}},\ \bibinfo {pages} {104426} (\bibinfo {year} {2021})}\BibitemShut {NoStop}%
\bibitem [{\citenamefont {Sale}\ \emph {et~al.}(2022)\citenamefont {Sale}, \citenamefont {Giansiracusa},\ and\ \citenamefont {Lucini}}]{PHPC2}%
  \BibitemOpen
  \bibfield  {author} {\bibinfo {author} {\bibfnamefont {N.}~\bibnamefont {Sale}}, \bibinfo {author} {\bibfnamefont {J.}~\bibnamefont {Giansiracusa}}, \ and\ \bibinfo {author} {\bibfnamefont {B.}~\bibnamefont {Lucini}},\ }\href {\doibase 10.1103/PhysRevE.105.024121} {\bibfield  {journal} {\bibinfo  {journal} {Phys. Rev. E}\ }\textbf {\bibinfo {volume} {105}},\ \bibinfo {pages} {024121} (\bibinfo {year} {2022})}\BibitemShut {NoStop}%
\bibitem [{\citenamefont {Carlsson}(2009{\natexlab{a}})}]{PHPC3}%
  \BibitemOpen
  \bibfield  {author} {\bibinfo {author} {\bibfnamefont {G.}~\bibnamefont {Carlsson}},\ }\href {\doibase 10.1090/S0273-0979-09-01249-X} {\bibfield  {journal} {\bibinfo  {journal} {Bull. Am. Math. Soc.}\ }\textbf {\bibinfo {volume} {46}},\ \bibinfo {pages} {255} (\bibinfo {year} {2009}{\natexlab{a}})}\BibitemShut {NoStop}%
\bibitem [{\citenamefont {Wulf}\ and\ \citenamefont {Muench}(2021)}]{TDAPC}%
  \BibitemOpen
  \bibfield  {author} {\bibinfo {author} {\bibfnamefont {J.~B.}\ \bibnamefont {Wulf}}\ and\ \bibinfo {author} {\bibfnamefont {I.}~\bibnamefont {Muench}},\ }\href {\doibase 10.1002/pamm.202000163} {\bibfield  {journal} {\bibinfo  {journal} {Proc. Appl. Math. Mech.}\ }\textbf {\bibinfo {volume} {20}},\ \bibinfo {pages} {e202000163} (\bibinfo {year} {2021})}\BibitemShut {NoStop}%
\bibitem [{\citenamefont {Zia}\ \emph {et~al.}(2024)\citenamefont {Zia}, \citenamefont {Khamis}, \citenamefont {Nichols} \emph {et~al.}}]{TML1}%
  \BibitemOpen
  \bibfield  {author} {\bibinfo {author} {\bibfnamefont {A.}~\bibnamefont {Zia}}, \bibinfo {author} {\bibfnamefont {A.}~\bibnamefont {Khamis}}, \bibinfo {author} {\bibfnamefont {J.}~\bibnamefont {Nichols}},  \emph {et~al.},\ }\href {\doibase 10.1007/s10462-024-10710-9} {\bibfield  {journal} {\bibinfo  {journal} {Artif. Intell. Rev.}\ }\textbf {\bibinfo {volume} {57}},\ \bibinfo {pages} {77} (\bibinfo {year} {2024})}\BibitemShut {NoStop}%
\bibitem [{\citenamefont {Leykam}\ and\ \citenamefont {Angelakis}(2023)}]{TML2}%
  \BibitemOpen
  \bibfield  {author} {\bibinfo {author} {\bibfnamefont {D.}~\bibnamefont {Leykam}}\ and\ \bibinfo {author} {\bibfnamefont {D.~G.}\ \bibnamefont {Angelakis}},\ }\href {\doibase 10.1080/23746149.2023.2202331} {\bibfield  {journal} {\bibinfo  {journal} {Adv. Phys. X}\ }\textbf {\bibinfo {volume} {8}},\ \bibinfo {pages} {2202331} (\bibinfo {year} {2023})}\BibitemShut {NoStop}%
\bibitem [{\citenamefont {Hensel}\ \emph {et~al.}(2021)\citenamefont {Hensel}, \citenamefont {Moor},\ and\ \citenamefont {Rieck}}]{TML3}%
  \BibitemOpen
  \bibfield  {author} {\bibinfo {author} {\bibfnamefont {F.}~\bibnamefont {Hensel}}, \bibinfo {author} {\bibfnamefont {M.}~\bibnamefont {Moor}}, \ and\ \bibinfo {author} {\bibfnamefont {B.}~\bibnamefont {Rieck}},\ }\href {\doibase 10.3389/frai.2021.681108} {\bibfield  {journal} {\bibinfo  {journal} {Front. Artif. Intell.}\ }\textbf {\bibinfo {volume} {4}},\ \bibinfo {pages} {681108} (\bibinfo {year} {2021})}\BibitemShut {NoStop}%
\bibitem [{\citenamefont {Bergomi}\ \emph {et~al.}(2019)\citenamefont {Bergomi}, \citenamefont {Frosini}, \citenamefont {Giorgi} \emph {et~al.}}]{TML}%
  \BibitemOpen
  \bibfield  {author} {\bibinfo {author} {\bibfnamefont {M.~G.}\ \bibnamefont {Bergomi}}, \bibinfo {author} {\bibfnamefont {P.}~\bibnamefont {Frosini}}, \bibinfo {author} {\bibfnamefont {D.}~\bibnamefont {Giorgi}},  \emph {et~al.},\ }\href {\doibase 10.1038/s42256-019-0087-3} {\bibfield  {journal} {\bibinfo  {journal} {Nat. Mach. Intell.}\ }\textbf {\bibinfo {volume} {1}},\ \bibinfo {pages} {423} (\bibinfo {year} {2019})}\BibitemShut {NoStop}%
\bibitem [{\citenamefont {Ghrist}(2008)}]{PH_features1}%
  \BibitemOpen
  \bibfield  {author} {\bibinfo {author} {\bibfnamefont {R.}~\bibnamefont {Ghrist}},\ }\href {\doibase 10.1090/S0273-0979-07-01191-3} {\bibfield  {journal} {\bibinfo  {journal} {Bull. Amer. Math. Soc.}\ }\textbf {\bibinfo {volume} {45}},\ \bibinfo {pages} {61} (\bibinfo {year} {2008})}\BibitemShut {NoStop}%
\bibitem [{\citenamefont {Zomorodian}\ and\ \citenamefont {Carlsson}(2005)}]{PH_features2}%
  \BibitemOpen
  \bibfield  {author} {\bibinfo {author} {\bibfnamefont {A.}~\bibnamefont {Zomorodian}}\ and\ \bibinfo {author} {\bibfnamefont {G.}~\bibnamefont {Carlsson}},\ }\href {\doibase 10.1007/s00454-004-1146-y} {\bibfield  {journal} {\bibinfo  {journal} {Discrete Comput. Geom.}\ }\textbf {\bibinfo {volume} {33}},\ \bibinfo {pages} {249} (\bibinfo {year} {2005})}\BibitemShut {NoStop}%
\bibitem [{\citenamefont {Carlsson}(2009{\natexlab{b}})}]{PH_features3}%
  \BibitemOpen
  \bibfield  {author} {\bibinfo {author} {\bibfnamefont {G.}~\bibnamefont {Carlsson}},\ }\href {\doibase 10.1090/S0273-0979-09-01249-X} {\bibfield  {journal} {\bibinfo  {journal} {Bull. Amer. Math. Soc.}\ }\textbf {\bibinfo {volume} {46}},\ \bibinfo {pages} {255} (\bibinfo {year} {2009}{\natexlab{b}})}\BibitemShut {NoStop}%
\bibitem [{\citenamefont {Chazal}\ and\ \citenamefont {Michel}(2021)}]{TDA_Introduction}%
  \BibitemOpen
  \bibfield  {author} {\bibinfo {author} {\bibfnamefont {F.}~\bibnamefont {Chazal}}\ and\ \bibinfo {author} {\bibfnamefont {B.}~\bibnamefont {Michel}},\ }\href {\doibase 10.3389/frai.2021.667963} {\bibfield  {journal} {\bibinfo  {journal} {Front. Artif. Intell.}\ }\textbf {\bibinfo {volume} {4}},\ \bibinfo {pages} {667963} (\bibinfo {year} {2021})}\BibitemShut {NoStop}%
\bibitem [{\citenamefont {Xu}\ \emph {et~al.}(2019)\citenamefont {Xu}, \citenamefont {Cisewski-Kehe}, \citenamefont {Green} \emph {et~al.}}]{applycosmic1}%
  \BibitemOpen
  \bibfield  {author} {\bibinfo {author} {\bibfnamefont {X.}~\bibnamefont {Xu}}, \bibinfo {author} {\bibfnamefont {J.}~\bibnamefont {Cisewski-Kehe}}, \bibinfo {author} {\bibfnamefont {S.~B.}\ \bibnamefont {Green}},  \emph {et~al.},\ }\href {\doibase 10.1016/j.ascom.2019.02.003} {\bibfield  {journal} {\bibinfo  {journal} {Astron. Comput.}\ }\textbf {\bibinfo {volume} {27}},\ \bibinfo {pages} {34} (\bibinfo {year} {2019})}\BibitemShut {NoStop}%
\bibitem [{\citenamefont {Biagetti}\ \emph {et~al.}(2021)\citenamefont {Biagetti}, \citenamefont {Cole},\ and\ \citenamefont {Shiu}}]{applycosmic2}%
  \BibitemOpen
  \bibfield  {author} {\bibinfo {author} {\bibfnamefont {M.}~\bibnamefont {Biagetti}}, \bibinfo {author} {\bibfnamefont {A.}~\bibnamefont {Cole}}, \ and\ \bibinfo {author} {\bibfnamefont {G.}~\bibnamefont {Shiu}},\ }\href {\doibase 10.1088/1475-7516/2021/04/061} {\bibfield  {journal} {\bibinfo  {journal} {J. Cosmol. Astropart. Phys.}\ }\textbf {\bibinfo {volume} {2021}},\ \bibinfo {pages} {061} (\bibinfo {year} {2021})}\BibitemShut {NoStop}%
\bibitem [{\citenamefont {Wilding}\ \emph {et~al.}(2021)\citenamefont {Wilding}, \citenamefont {Nevenzeel}, \citenamefont {van~de Weygaert} \emph {et~al.}}]{applycosmic3}%
  \BibitemOpen
  \bibfield  {author} {\bibinfo {author} {\bibfnamefont {G.}~\bibnamefont {Wilding}}, \bibinfo {author} {\bibfnamefont {K.}~\bibnamefont {Nevenzeel}}, \bibinfo {author} {\bibfnamefont {R.}~\bibnamefont {van~de Weygaert}},  \emph {et~al.},\ }\href {\doibase 10.1093/mnras/stab2326} {\bibfield  {journal} {\bibinfo  {journal} {MNRAS}\ }\textbf {\bibinfo {volume} {507}},\ \bibinfo {pages} {2968} (\bibinfo {year} {2021})}\BibitemShut {NoStop}%
\bibitem [{\citenamefont {Spitz}\ \emph {et~al.}(2023)\citenamefont {Spitz}, \citenamefont {Boguslavski},\ and\ \citenamefont {Berges}}]{pointandtopo}%
  \BibitemOpen
  \bibfield  {author} {\bibinfo {author} {\bibfnamefont {D.}~\bibnamefont {Spitz}}, \bibinfo {author} {\bibfnamefont {K.}~\bibnamefont {Boguslavski}}, \ and\ \bibinfo {author} {\bibfnamefont {J.}~\bibnamefont {Berges}},\ }\href {\doibase 10.1103/PhysRevD.108.056016} {\bibfield  {journal} {\bibinfo  {journal} {Phys. Rev. D}\ }\textbf {\bibinfo {volume} {108}},\ \bibinfo {pages} {056016} (\bibinfo {year} {2023})}\BibitemShut {NoStop}%
\bibitem [{\citenamefont {Bialas}\ and\ \citenamefont {Peschanski}(1986)}]{intermittency_and_fractal_dimension1}%
  \BibitemOpen
  \bibfield  {author} {\bibinfo {author} {\bibfnamefont {A.}~\bibnamefont {Bialas}}\ and\ \bibinfo {author} {\bibfnamefont {R.}~\bibnamefont {Peschanski}},\ }\href {\doibase 10.1016/0550-3213(86)90386-X} {\bibfield  {journal} {\bibinfo  {journal} {Nucl. Phys. B}\ }\textbf {\bibinfo {volume} {273}},\ \bibinfo {pages} {703} (\bibinfo {year} {1986})}\BibitemShut {NoStop}%
\bibitem [{\citenamefont {Bialas}\ and\ \citenamefont {Peschanski}(1988)}]{intermittency_and_fractal_dimension2}%
  \BibitemOpen
  \bibfield  {author} {\bibinfo {author} {\bibfnamefont {A.}~\bibnamefont {Bialas}}\ and\ \bibinfo {author} {\bibfnamefont {R.}~\bibnamefont {Peschanski}},\ }\href {\doibase 10.1016/0550-3213(88)90131-9} {\bibfield  {journal} {\bibinfo  {journal} {Nucl. Phys. B}\ }\textbf {\bibinfo {volume} {308}},\ \bibinfo {pages} {857} (\bibinfo {year} {1988})}\BibitemShut {NoStop}%
\bibitem [{\citenamefont {De~Wolf}\ \emph {et~al.}(1996)\citenamefont {De~Wolf}, \citenamefont {Dremin},\ and\ \citenamefont {Kittel}}]{intermittency_and_fractal_dimension3}%
  \BibitemOpen
  \bibfield  {author} {\bibinfo {author} {\bibfnamefont {E.~A.}\ \bibnamefont {De~Wolf}}, \bibinfo {author} {\bibfnamefont {I.}~\bibnamefont {Dremin}}, \ and\ \bibinfo {author} {\bibfnamefont {W.}~\bibnamefont {Kittel}},\ }\href {\doibase 10.1016/0370-1573(95)00069-0} {\bibfield  {journal} {\bibinfo  {journal} {Phys. Rep.}\ }\textbf {\bibinfo {volume} {270}},\ \bibinfo {pages} {1} (\bibinfo {year} {1996})}\BibitemShut {NoStop}%
\bibitem [{\citenamefont {Antoniou}\ \emph {et~al.}(2005)\citenamefont {Antoniou}, \citenamefont {Contoyiannis}, \citenamefont {Diakonos} \emph {et~al.}}]{intermittencyindex}%
  \BibitemOpen
  \bibfield  {author} {\bibinfo {author} {\bibfnamefont {N.~G.}\ \bibnamefont {Antoniou}}, \bibinfo {author} {\bibfnamefont {Y.~F.}\ \bibnamefont {Contoyiannis}}, \bibinfo {author} {\bibfnamefont {F.~K.}\ \bibnamefont {Diakonos}},  \emph {et~al.},\ }\href {\doibase 10.1016/j.nuclphysa.2005.07.003} {\bibfield  {journal} {\bibinfo  {journal} {Nucl. Phys. A}\ }\textbf {\bibinfo {volume} {761}},\ \bibinfo {pages} {149} (\bibinfo {year} {2005})}\BibitemShut {NoStop}%
\bibitem [{\citenamefont {Antoniou}\ \emph {et~al.}(2007)\citenamefont {Antoniou}, \citenamefont {Diakonos},\ and\ \citenamefont {Saridakis}}]{intermittencyindex-2}%
  \BibitemOpen
  \bibfield  {author} {\bibinfo {author} {\bibfnamefont {N.~G.}\ \bibnamefont {Antoniou}}, \bibinfo {author} {\bibfnamefont {F.~K.}\ \bibnamefont {Diakonos}}, \ and\ \bibinfo {author} {\bibfnamefont {E.~N.}\ \bibnamefont {Saridakis}},\ }\href {\doibase 10.1016/j.nuclphysa.2006.12.006} {\bibfield  {journal} {\bibinfo  {journal} {Nucl. Phys. A}\ }\textbf {\bibinfo {volume} {784}},\ \bibinfo {pages} {536} (\bibinfo {year} {2007})}\BibitemShut {NoStop}%
\bibitem [{\citenamefont {Antoniou}\ \emph {et~al.}(2008)\citenamefont {Antoniou}, \citenamefont {Diakonos},\ and\ \citenamefont {Saridakis}}]{intermittencyindex-3}%
  \BibitemOpen
  \bibfield  {author} {\bibinfo {author} {\bibfnamefont {N.~G.}\ \bibnamefont {Antoniou}}, \bibinfo {author} {\bibfnamefont {F.~K.}\ \bibnamefont {Diakonos}}, \ and\ \bibinfo {author} {\bibfnamefont {E.~N.}\ \bibnamefont {Saridakis}},\ }\href {\doibase 10.1103/PhysRevC.78.024908} {\bibfield  {journal} {\bibinfo  {journal} {Phys. Rev. C}\ }\textbf {\bibinfo {volume} {78}},\ \bibinfo {pages} {024908} (\bibinfo {year} {2008})}\BibitemShut {NoStop}%
\bibitem [{\citenamefont {Anticic}\ \emph {et~al.}(2010)\citenamefont {Anticic}, \citenamefont {Baatar}, \citenamefont {Barna} \emph {et~al.}}]{intermittencyindex-4}%
  \BibitemOpen
  \bibfield  {author} {\bibinfo {author} {\bibfnamefont {T.}~\bibnamefont {Anticic}}, \bibinfo {author} {\bibfnamefont {B.}~\bibnamefont {Baatar}}, \bibinfo {author} {\bibfnamefont {D.}~\bibnamefont {Barna}},  \emph {et~al.},\ }\href {\doibase 10.1103/PhysRevC.81.064907} {\bibfield  {journal} {\bibinfo  {journal} {Phys. Rev. C}\ }\textbf {\bibinfo {volume} {81}},\ \bibinfo {pages} {064907} (\bibinfo {year} {2010})}\BibitemShut {NoStop}%
\bibitem [{\citenamefont {Alemany}\ and\ \citenamefont {Zanette}(1994)}]{Levy_random}%
  \BibitemOpen
  \bibfield  {author} {\bibinfo {author} {\bibfnamefont {P.~A.}\ \bibnamefont {Alemany}}\ and\ \bibinfo {author} {\bibfnamefont {D.~H.}\ \bibnamefont {Zanette}},\ }\href {\doibase 10.1103/PhysRevE.49.R956} {\bibfield  {journal} {\bibinfo  {journal} {Phys. Rev. E}\ }\textbf {\bibinfo {volume} {49}},\ \bibinfo {pages} {R956} (\bibinfo {year} {1994})}\BibitemShut {NoStop}%
\bibitem [{\citenamefont {Bleicher}\ \emph {et~al.}(1999)\citenamefont {Bleicher}, \citenamefont {Zabrodin}, \citenamefont {Spieles} \emph {et~al.}}]{UrQMD1}%
  \BibitemOpen
  \bibfield  {author} {\bibinfo {author} {\bibfnamefont {M.}~\bibnamefont {Bleicher}}, \bibinfo {author} {\bibfnamefont {E.}~\bibnamefont {Zabrodin}}, \bibinfo {author} {\bibfnamefont {C.}~\bibnamefont {Spieles}},  \emph {et~al.},\ }\href {\doibase 10.1088/0954-3899/25/9/308} {\bibfield  {journal} {\bibinfo  {journal} {J. Phys. G Nucl. Partic.}\ }\textbf {\bibinfo {volume} {25}},\ \bibinfo {pages} {1859} (\bibinfo {year} {1999})}\BibitemShut {NoStop}%
\bibitem [{\citenamefont {Bass}\ \emph {et~al.}(1998)\citenamefont {Bass}, \citenamefont {Belkacem}, \citenamefont {Bleicher} \emph {et~al.}}]{UrQMD2}%
  \BibitemOpen
  \bibfield  {author} {\bibinfo {author} {\bibfnamefont {S.~A.}\ \bibnamefont {Bass}}, \bibinfo {author} {\bibfnamefont {M.}~\bibnamefont {Belkacem}}, \bibinfo {author} {\bibfnamefont {M.}~\bibnamefont {Bleicher}},  \emph {et~al.},\ }\href {\doibase 10.1016/S0146-6410(98)00058-1} {\bibfield  {journal} {\bibinfo  {journal} {Prog. Part. Nucl. Phys.}\ }\textbf {\bibinfo {volume} {41}},\ \bibinfo {pages} {255} (\bibinfo {year} {1998})}\BibitemShut {NoStop}%
\bibitem [{\citenamefont {Petersen}\ \emph {et~al.}(2008)\citenamefont {Petersen}, \citenamefont {Steinheimer}, \citenamefont {Burau} \emph {et~al.}}]{UrQMD3}%
  \BibitemOpen
  \bibfield  {author} {\bibinfo {author} {\bibfnamefont {H.}~\bibnamefont {Petersen}}, \bibinfo {author} {\bibfnamefont {J.}~\bibnamefont {Steinheimer}}, \bibinfo {author} {\bibfnamefont {G.}~\bibnamefont {Burau}},  \emph {et~al.},\ }\href {\doibase 10.1103/PhysRevC.78.044901} {\bibfield  {journal} {\bibinfo  {journal} {Phys. Rev. C}\ }\textbf {\bibinfo {volume} {78}},\ \bibinfo {pages} {044901} (\bibinfo {year} {2008})}\BibitemShut {NoStop}%
\bibitem [{\citenamefont {Knospe}\ \emph {et~al.}(2016)\citenamefont {Knospe}, \citenamefont {Markert}, \citenamefont {Werner} \emph {et~al.}}]{UrQMD4}%
  \BibitemOpen
  \bibfield  {author} {\bibinfo {author} {\bibfnamefont {A.~G.}\ \bibnamefont {Knospe}}, \bibinfo {author} {\bibfnamefont {C.}~\bibnamefont {Markert}}, \bibinfo {author} {\bibfnamefont {K.}~\bibnamefont {Werner}},  \emph {et~al.},\ }\href {\doibase 10.1103/PhysRevC.93.014911} {\bibfield  {journal} {\bibinfo  {journal} {Phys. Rev. C}\ }\textbf {\bibinfo {volume} {93}},\ \bibinfo {pages} {014911} (\bibinfo {year} {2016})}\BibitemShut {NoStop}%
\bibitem [{\citenamefont {Wu}(2022)}]{STAR1}%
  \BibitemOpen
  \bibfield  {author} {\bibinfo {author} {\bibfnamefont {J.}~\bibnamefont {Wu}} (\bibinfo {collaboration} {for the STAR Collaboration}),\ }\href {\doibase 10.21468/SciPostPhysProc.10.041} {\bibfield  {journal} {\bibinfo  {journal} {SciPost Phys. Proc}\ }\textbf {\bibinfo {volume} {10}},\ \bibinfo {pages} {041} (\bibinfo {year} {2022})}\BibitemShut {NoStop}%
\bibitem [{\citenamefont {Efron}\ and\ \citenamefont {Tibshirani}(1986)}]{RefBootstrap}%
  \BibitemOpen
  \bibfield  {author} {\bibinfo {author} {\bibfnamefont {B.}~\bibnamefont {Efron}}\ and\ \bibinfo {author} {\bibfnamefont {R.}~\bibnamefont {Tibshirani}},\ }\href {http://dml.mathdoc.fr/item/1177013815} {\bibfield  {journal} {\bibinfo  {journal} {Statist. Sci.}\ }\textbf {\bibinfo {volume} {1}},\ \bibinfo {pages} {54} (\bibinfo {year} {1986})}\BibitemShut {NoStop}%
\bibitem [{\citenamefont {Kaczynski}\ \emph {et~al.}(2004{\natexlab{a}})\citenamefont {Kaczynski}, \citenamefont {Mischaikow},\ and\ \citenamefont {Mrozek}}]{Topology}%
  \BibitemOpen
  \bibfield  {author} {\bibinfo {author} {\bibfnamefont {T.}~\bibnamefont {Kaczynski}}, \bibinfo {author} {\bibfnamefont {K.}~\bibnamefont {Mischaikow}}, \ and\ \bibinfo {author} {\bibfnamefont {M.}~\bibnamefont {Mrozek}},\ }\href {\doibase 10.1007/0-387-21597-2_4} {\bibfield  {journal} {\bibinfo  {journal} {Comput. Homol.}\ }\textbf {\bibinfo {volume} {1}},\ \bibinfo {pages} {143} (\bibinfo {year} {2004}{\natexlab{a}})}\BibitemShut {NoStop}%
\bibitem [{\citenamefont {Munkres}(2018)}]{cal_betti}%
  \BibitemOpen
  \bibfield  {author} {\bibinfo {author} {\bibfnamefont {J.~R.}\ \bibnamefont {Munkres}},\ }\href {\doibase 10.1201/9780429493911} {\emph {\bibinfo {title} {Elements of algebraic topology}}}\ (\bibinfo  {publisher} {CRC press},\ \bibinfo {year} {2018})\BibitemShut {NoStop}%
\bibitem [{\citenamefont {Otter}\ \emph {et~al.}(2017)\citenamefont {Otter}, \citenamefont {Porter}, \citenamefont {Tillmann} \emph {et~al.}}]{DTFE2}%
  \BibitemOpen
  \bibfield  {author} {\bibinfo {author} {\bibfnamefont {N.}~\bibnamefont {Otter}}, \bibinfo {author} {\bibfnamefont {M.~A.}\ \bibnamefont {Porter}}, \bibinfo {author} {\bibfnamefont {U.}~\bibnamefont {Tillmann}},  \emph {et~al.},\ }\href {\doibase 10.1140/epjds/s13688-017-0109-5} {\bibfield  {journal} {\bibinfo  {journal} {EPJ Data Sci.}\ }\textbf {\bibinfo {volume} {6}},\ \bibinfo {pages} {1} (\bibinfo {year} {2017})}\BibitemShut {NoStop}%
\bibitem [{\citenamefont {Reitberger}(2002)}]{VRComplex}%
  \BibitemOpen
  \bibfield  {author} {\bibinfo {author} {\bibfnamefont {H.}~\bibnamefont {Reitberger}},\ }\href {https://www.ams.org/notices/200210/fea-vietoris.pdf} {\bibfield  {journal} {\bibinfo  {journal} {Not. Am. Math. Soc.}\ }\textbf {\bibinfo {volume} {49}},\ \bibinfo {pages} {1232} (\bibinfo {year} {2002})}\BibitemShut {NoStop}%
\bibitem [{\citenamefont {Edelsbrunner}\ \emph {et~al.}(1983)\citenamefont {Edelsbrunner}, \citenamefont {Kirkpatrick},\ and\ \citenamefont {Seidel}}]{AlphaComplex-1}%
  \BibitemOpen
  \bibfield  {author} {\bibinfo {author} {\bibfnamefont {H.}~\bibnamefont {Edelsbrunner}}, \bibinfo {author} {\bibfnamefont {D.}~\bibnamefont {Kirkpatrick}}, \ and\ \bibinfo {author} {\bibfnamefont {R.}~\bibnamefont {Seidel}},\ }\href {\doibase 10.1109/TIT.1983.1056714} {\bibfield  {journal} {\bibinfo  {journal} {IEEE Trans. Inf. Theory}\ }\textbf {\bibinfo {volume} {29}},\ \bibinfo {pages} {551} (\bibinfo {year} {1983})}\BibitemShut {NoStop}%
\bibitem [{\citenamefont {Edelsbrunner}\ and\ \citenamefont {M{\"u}cke}(1994)}]{AlphaComplex-2}%
  \BibitemOpen
  \bibfield  {author} {\bibinfo {author} {\bibfnamefont {H.}~\bibnamefont {Edelsbrunner}}\ and\ \bibinfo {author} {\bibfnamefont {E.~P.}\ \bibnamefont {M{\"u}cke}},\ }\href {\doibase 10.1145/174462.156635} {\bibfield  {journal} {\bibinfo  {journal} {ACM Trans. Graphics}\ }\textbf {\bibinfo {volume} {13}},\ \bibinfo {pages} {43} (\bibinfo {year} {1994})}\BibitemShut {NoStop}%
\bibitem [{\citenamefont {Edelsbrunner}\ and\ \citenamefont {Harer}(2022)}]{TopologicalComplexes}%
  \BibitemOpen
  \bibfield  {author} {\bibinfo {author} {\bibfnamefont {H.}~\bibnamefont {Edelsbrunner}}\ and\ \bibinfo {author} {\bibfnamefont {J.}~\bibnamefont {Harer}},\ }\href {\doibase 10.1007/978-3-540-33259-6_7} {\emph {\bibinfo {title} {Computational topology: an introduction}}}\ (\bibinfo  {publisher} {American Mathematical Society},\ \bibinfo {year} {2022})\BibitemShut {NoStop}%
\bibitem [{\citenamefont {Delaunay}(1934)}]{delaunay1}%
  \BibitemOpen
  \bibfield  {author} {\bibinfo {author} {\bibfnamefont {B.}~\bibnamefont {Delaunay}},\ }\href {https://www.mathnet.ru/links/4dc69632645526bd521de4d493371ca0/im4937.pdf} {\bibfield  {journal} {\bibinfo  {journal} {Izvestia Akademii Nauk SSSR: Otdelenie Matematicheskikh i Estestvennykh Nauk}\ }\textbf {\bibinfo {volume} {7}},\ \bibinfo {pages} {793} (\bibinfo {year} {1934})}\BibitemShut {NoStop}%
\bibitem [{\citenamefont {Lee}\ and\ \citenamefont {Schachter}(1980)}]{delaunay2}%
  \BibitemOpen
  \bibfield  {author} {\bibinfo {author} {\bibfnamefont {D.~T.}\ \bibnamefont {Lee}}\ and\ \bibinfo {author} {\bibfnamefont {B.~J.}\ \bibnamefont {Schachter}},\ }\href {\doibase 10.1007/BF00977785} {\bibfield  {journal} {\bibinfo  {journal} {Int. j. comput. inf. sci}\ }\textbf {\bibinfo {volume} {9}},\ \bibinfo {pages} {219} (\bibinfo {year} {1980})}\BibitemShut {NoStop}%
\bibitem [{\citenamefont {Frisken}\ \emph {et~al.}(2000)\citenamefont {Frisken}, \citenamefont {Perry}, \citenamefont {Rockwood} \emph {et~al.}}]{distanceField}%
  \BibitemOpen
  \bibfield  {author} {\bibinfo {author} {\bibfnamefont {S.~F.}\ \bibnamefont {Frisken}}, \bibinfo {author} {\bibfnamefont {R.~N.}\ \bibnamefont {Perry}}, \bibinfo {author} {\bibfnamefont {A.~P.}\ \bibnamefont {Rockwood}},  \emph {et~al.},\ }in\ \href {\doibase 10.1145/344779.344899} {\emph {\bibinfo {booktitle} {Proceedings of the 27th annual conference on Computer graphics and interactive techniques}}}\ (\bibinfo {year} {2000})\ pp.\ \bibinfo {pages} {249--254}\BibitemShut {NoStop}%
\bibitem [{\citenamefont {Edelsbrunner}\ and\ \citenamefont {Harer}(2008)}]{SublevelFiltration-1}%
  \BibitemOpen
  \bibfield  {author} {\bibinfo {author} {\bibfnamefont {H.}~\bibnamefont {Edelsbrunner}}\ and\ \bibinfo {author} {\bibfnamefont {J.}~\bibnamefont {Harer}},\ }\href {\doibase 10.1090/conm/453/08802} {\bibfield  {journal} {\bibinfo  {journal} {Contemporary mathematics}\ }\textbf {\bibinfo {volume} {453}},\ \bibinfo {pages} {257} (\bibinfo {year} {2008})}\BibitemShut {NoStop}%
\bibitem [{\citenamefont {Kaczynski}\ \emph {et~al.}(2004{\natexlab{b}})\citenamefont {Kaczynski}, \citenamefont {Mischaikow},\ and\ \citenamefont {Mrozek}}]{SublevelFiltration-2}%
  \BibitemOpen
  \bibfield  {author} {\bibinfo {author} {\bibfnamefont {T.}~\bibnamefont {Kaczynski}}, \bibinfo {author} {\bibfnamefont {K.~M.}\ \bibnamefont {Mischaikow}}, \ and\ \bibinfo {author} {\bibfnamefont {M.}~\bibnamefont {Mrozek}},\ }\href {\doibase 10.1007/b97315} {\emph {\bibinfo {title} {Computational Homology}}},\ \bibinfo {series} {Applied Mathematical Sciences}, Vol.\ \bibinfo {volume} {157}\ (\bibinfo  {publisher} {Springer-Verlag},\ \bibinfo {address} {New York},\ \bibinfo {year} {2004})\BibitemShut {NoStop}%
\bibitem [{\citenamefont {Dummit}\ and\ \citenamefont {Foote}(2003)}]{dummit2003abstract}%
  \BibitemOpen
  \bibfield  {author} {\bibinfo {author} {\bibfnamefont {D.}~\bibnamefont {Dummit}}\ and\ \bibinfo {author} {\bibfnamefont {R.}~\bibnamefont {Foote}},\ }\href {https://books.google.com.sg/books?id=KJDBQgAACAAJ} {\emph {\bibinfo {title} {Abstract Algebra}}}\ (\bibinfo  {publisher} {Wiley},\ \bibinfo {year} {2003})\BibitemShut {NoStop}%
\bibitem [{\citenamefont {Greenberg}\ and\ \citenamefont {Harper}(2019)}]{greenberg2019algebraic}%
  \BibitemOpen
  \bibfield  {author} {\bibinfo {author} {\bibfnamefont {M.}~\bibnamefont {Greenberg}}\ and\ \bibinfo {author} {\bibfnamefont {J.}~\bibnamefont {Harper}},\ }\href {https://books.google.com.sg/books?id=GBQVyAEACAAJ} {\emph {\bibinfo {title} {Algebraic Topology: A First Course}}},\ Mathematics Lecture Note Series\ (\bibinfo  {publisher} {Taylor \& Francis Group},\ \bibinfo {year} {2019})\BibitemShut {NoStop}%
\bibitem [{\citenamefont {Bubenik}(2015)}]{PH_features4}%
  \BibitemOpen
  \bibfield  {author} {\bibinfo {author} {\bibfnamefont {P.}~\bibnamefont {Bubenik}},\ }\href {http://jmlr.org/papers/v16/bubenik15a.html} {\bibfield  {journal} {\bibinfo  {journal} {J MACH LEARN RES}\ }\textbf {\bibinfo {volume} {16}},\ \bibinfo {pages} {77} (\bibinfo {year} {2015})}\BibitemShut {NoStop}%
\bibitem [{\citenamefont {Adams}\ \emph {et~al.}(2017)\citenamefont {Adams}, \citenamefont {Emerson}, \citenamefont {Kirby} \emph {et~al.}}]{PH_features5}%
  \BibitemOpen
  \bibfield  {author} {\bibinfo {author} {\bibfnamefont {H.}~\bibnamefont {Adams}}, \bibinfo {author} {\bibfnamefont {T.}~\bibnamefont {Emerson}}, \bibinfo {author} {\bibfnamefont {M.}~\bibnamefont {Kirby}},  \emph {et~al.},\ }\href {http://jmlr.org/papers/v18/16-337.html} {\bibfield  {journal} {\bibinfo  {journal} {J MACH LEARN RES}\ }\textbf {\bibinfo {volume} {18}},\ \bibinfo {pages} {1} (\bibinfo {year} {2017})}\BibitemShut {NoStop}%
\bibitem [{\citenamefont {Chung}\ and\ \citenamefont {Lawson}(2022)}]{PH_features6}%
  \BibitemOpen
  \bibfield  {author} {\bibinfo {author} {\bibfnamefont {Y.-M.}\ \bibnamefont {Chung}}\ and\ \bibinfo {author} {\bibfnamefont {A.}~\bibnamefont {Lawson}},\ }\href {\doibase https://doi.org/10.1007/s10444-021-09893-4} {\bibfield  {journal} {\bibinfo  {journal} {Adv. Comput. Math.}\ }\textbf {\bibinfo {volume} {48}},\ \bibinfo {pages} {6} (\bibinfo {year} {2022})}\BibitemShut {NoStop}%
\bibitem [{\citenamefont {Charles}\ \emph {et~al.}(2017)\citenamefont {Charles}, \citenamefont {Su}, \citenamefont {Kaichun} \emph {et~al.}}]{PointNet}%
  \BibitemOpen
  \bibfield  {author} {\bibinfo {author} {\bibfnamefont {R.~Q.}\ \bibnamefont {Charles}}, \bibinfo {author} {\bibfnamefont {H.}~\bibnamefont {Su}}, \bibinfo {author} {\bibfnamefont {M.}~\bibnamefont {Kaichun}},  \emph {et~al.},\ }in\ \href {\doibase 10.1109/CVPR.2017.16} {\emph {\bibinfo {booktitle} {2017 IEEE Conference on Computer Vision and Pattern Recognition (CVPR)}}}\ (\bibinfo {year} {2017})\ pp.\ \bibinfo {pages} {77--85}\BibitemShut {NoStop}%
\bibitem [{\citenamefont {Zhang}\ \emph {et~al.}(2023)\citenamefont {Zhang}, \citenamefont {Wang}, \citenamefont {Wang} \emph {et~al.}}]{PointNN}%
  \BibitemOpen
  \bibfield  {author} {\bibinfo {author} {\bibfnamefont {R.}~\bibnamefont {Zhang}}, \bibinfo {author} {\bibfnamefont {L.}~\bibnamefont {Wang}}, \bibinfo {author} {\bibfnamefont {Y.}~\bibnamefont {Wang}},  \emph {et~al.},\ }in\ \href {\doibase 10.1109/CVPR52729.2023.00517} {\emph {\bibinfo {booktitle} {Proceedings of the IEEE/CVF Conference on Computer Vision and Pattern Recognition (CVPR)}}}\ (\bibinfo {year} {2023})\ pp.\ \bibinfo {pages} {5344--5353}\BibitemShut {NoStop}%
\end{thebibliography}%
\end{document}